\documentclass{aa}

\usepackage{savesym}
\usepackage{amsmath}
\savesymbol{iint}
\savesymbol{iiint}
\usepackage{txfonts}
\restoresymbol{TXF}{iint}
\restoresymbol{TXF}{iiint}
\usepackage{natbib}
\bibpunct{(}{)}{;}{a}{}{,}
\usepackage{graphicx}

\begin{document} 

\title{Wave modes excited by photospheric p-modes and mode conversion in a multi-loop system}

\institute{Centre for mathematical Plasma Astrophysics (CmPA), KU Leuven \\ Celestijnenlaan 200B bus 2400 \\
3001 Leuven, Belgium \\
\email{juliamaria.riedl@kuleuven.be}} \label{inst}

\author{J.M. Riedl \ref{inst}
\and T. Van Doorsselaere \ref{inst}
\and I. C. Santamaria \ref{inst}} 

\date{Received 4 March 2019; 
Accepted 22 April 2019}

\abstract{Waves are ubiquitous in the solar corona and there are indications that they are excited by photospheric p-modes. However, it is unclear how p-modes in coronal loops are converted to sausage modes and transverse (kink) modes, which are observed in the corona.} {We aim to investigate how those wave modes are excited in the lower corona by photospheric acoustic waves.} {We built 3D magnetohydrostatic loop systems with multiple inclinations spanning from the photosphere to the lower corona. We then simulated these atmospheres with the MANCHA code, in which we perturb the equilibrium with a p-mode driver at the bottom of the domain. By splitting the velocity perturbation into components longitudinal, normal, and azimuthal to the magnetic flux surfaces we can study wave behavior.} {In vertical flux tubes, we find that deformed fast sausage surface waves and slow sausage body waves are excited. In inclined flux tubes fast kink surface waves, slow sausage body waves, and either a fast sausage surface wave or a plane wave are excited. In addition, we calculate a wave conversion factor ($0 \le C \le1$) from acoustic to magnetic wave behavior by taking the ratio of the mean magnetic energy flux to the sum of the mean magnetic and acoustic energy flux and compare it to a commonly used theoretical conversion factor. We find that between magnetic field inclinations of 10$\degr$ to 30$\degr$ those two methods lie within 40\%. For smaller inclinations the absolute deviation is smaller than 0.1. }{}


\maketitle

\section{Introduction} \label{sec:intro}

Ever since the discovery of the high temperatures of the solar corona by \cite{edlen_1943}, scientists have been trying to find an explanation for this phenomenon. One possible coronal heating mechanism is alternating current (AC) heating, in which magnetic energy is dissipated by waves \citep{aschwanden, priest_2014, parnell_demoortel_2012, arregui_2017}. AC heating mechanisms were widely ignored for a long time until it was observed that waves are indeed ubiquitous in the solar corona \citep{de_pontieu_etal_2007, tomczyk_etal_2007, krishna_etal_2012, morton_etal_2012, nistico_etal_2013}. 

\cite{tomczyk_mcintosh_2009} observed waves with the Coronal Multi-channel Polarimeter (CoMP) instrument and found that the spectrum of velocity perturbations peaks at the same frequency ($\sim$ 3 mHz) as solar p-modes. \cite{morton_etal_2016} and \cite{morton_etal_2019} confirmed that the power enhancement at this frequency is a global phenomenon. Therefore, it seems likely that the ubiquitous waves are at least partially driven by p-modes. However, it is not yet clearly understood how p-modes propagate into the higher atmosphere. Observations tracing p-modes in the chromosphere and above were done by, for example, \cite{centeno_etal_2006}, \cite{marsh_walsh_2006}, \cite{de_wijn_etal_2009}, \cite{prasad_etal_2015}, 
and \cite{zhao_etal_2016}, while numerical modeling was done by, for example, \cite{khomenko_etal_2008}, \cite{fedun_etal_2011}, \cite{santamaria_etal_2015}, and \cite{griffiths_2018}. \cite{de_pontieu_etal_2004} and \cite{de_pontieu_etal_2005} found that, although p-modes with periods above the cutoff period are usually evanescent in the chromosphere, these p-modes can propagate upward along inclined magnetic flux tubes because the effective cutoff period is increased in a non-vertical magnetic field. 

It was shown by \cite{bogdan_etal_1996}, \cite{hindman_jain_2008}, and \cite{gascoyne_etal_2014} that p-modes lose energy to magnetic tube waves, such as sausage waves and kink waves; the propagation of these excited waves into the solar atmosphere was not considered in those studies. Observations show that tube waves are indeed excited in the solar atmosphere. While propagating kink waves have already been observed many years ago \citep{verwichte_etal_2005}, \cite{grant_etal_2015} were more recently able to observe sausage modes propagating from the photosphere to the transition region for the first time. These authors find surprisingly strong damping for those sausage waves; this is not well understood and therefore indicates again our limited knowledge of wave propagation in that region.

We therefore stress the importance of understanding the propagation of p-modes through the chromosphere, as they might be the source of decayless transverse waves. Those waves could be connected to coronal heating \citep[e.g.,][]{karampelas_etal_2017}. They have been observed by \cite{wang_etal_2012}, \cite{nistico_etal_2013}, and \cite{anfinogentov_etal_2013}, for example. Transverse waves are currently modeled in the corona as loops with a horizontal driver in one or both footpoints \citep{karampelas_vandoorsselaere_2018,guo_etal_2019,karampelas_etal_2019, pagano_de_moortel_2019}, but from where those horizontal plasma movements originate is usually not discussed. One possible mechanism could be a self-oscillatory process due to the interaction of the loops with quasi-steady flows, as discussed by \cite{nakariakov_etal_2016}. Another possibility, as mentioned above, could be photospheric p-modes, which are converted to kink waves. 

In this work, we study the conversion of photospheric p-modes to sausage and kink waves in an atmosphere that is gravitationally stratified and has additionally structuring
perpendicular to the magnetic field. The model we use is in magnetohydrostatic (MHS) equilibrium and contains four loops with different inclinations ranging from the photosphere to the lower corona. We perturb the equilibrium at the bottom with a p-mode driver in the form of an analytic solution for gravity-acoustic waves and simulate the propagation of the waves. The waves interact with the cylindrical structure of our model atmosphere and tube modes are excited. The MHS model is described in Sect. \ref{sec:model}, while we explain the numerical setup in Sect. \ref{sec:numerics}. We present assisting methods for data interpretation in Sect. \ref{sec:methods} and present and discuss our results, including what kind of wave modes are excited in the corona and how their basic properties change, in Sect. \ref{sec:results}. Finally, we present the summary and conclusions of our work in Sect. \ref{sec:summary&conclusions}.

\section{Magnetohydrostatic equilibrium model} \label{sec:model}

We built a 3D MHS equilibrium atmosphere from the photosphere to the lower corona, which has to fulfill the condition
\begin{equation} \label{eq:mhs_equilibrium}
  \vec{\nabla} p_0 - \frac{1}{\mu_0} \left( \vec{\nabla} \times \vec{B_0} \right) \times \vec{B_0}- \rho_0 \vec{g}=0,
\end{equation}
where $p_0$, $\vec{B_0}$, and $\rho_0$ are the equilibrium pressure, equilibrium magnetic field, and equilibrium density, respectively, and $\mu_0$ is the permeability of vacuum. For our models, the gravity vector $\vec{g}=(0,0,-g)$ is constant and points to the negative $z$-direction. In addition, we built the atmosphere as periodic in the horizontal directions. The domain has a size of $n_x \times n_y \times n_z =$ 140$\times$140$\times$840 points with a resolution of $\Delta x \times \Delta y \times \Delta z \approx $ 14.3$\times$14.3$\times$6.0 km, which results in a domain with the approximate measurements of 2$\times$2$\times$5 Mm. The bottom seven planes of grid cells of the domain are located below the photosphere ($z=0$) to make space for the driver (see Sect. \ref{sec:numerics}).

\begin{figure*}
\includegraphics[width=0.33\textwidth]{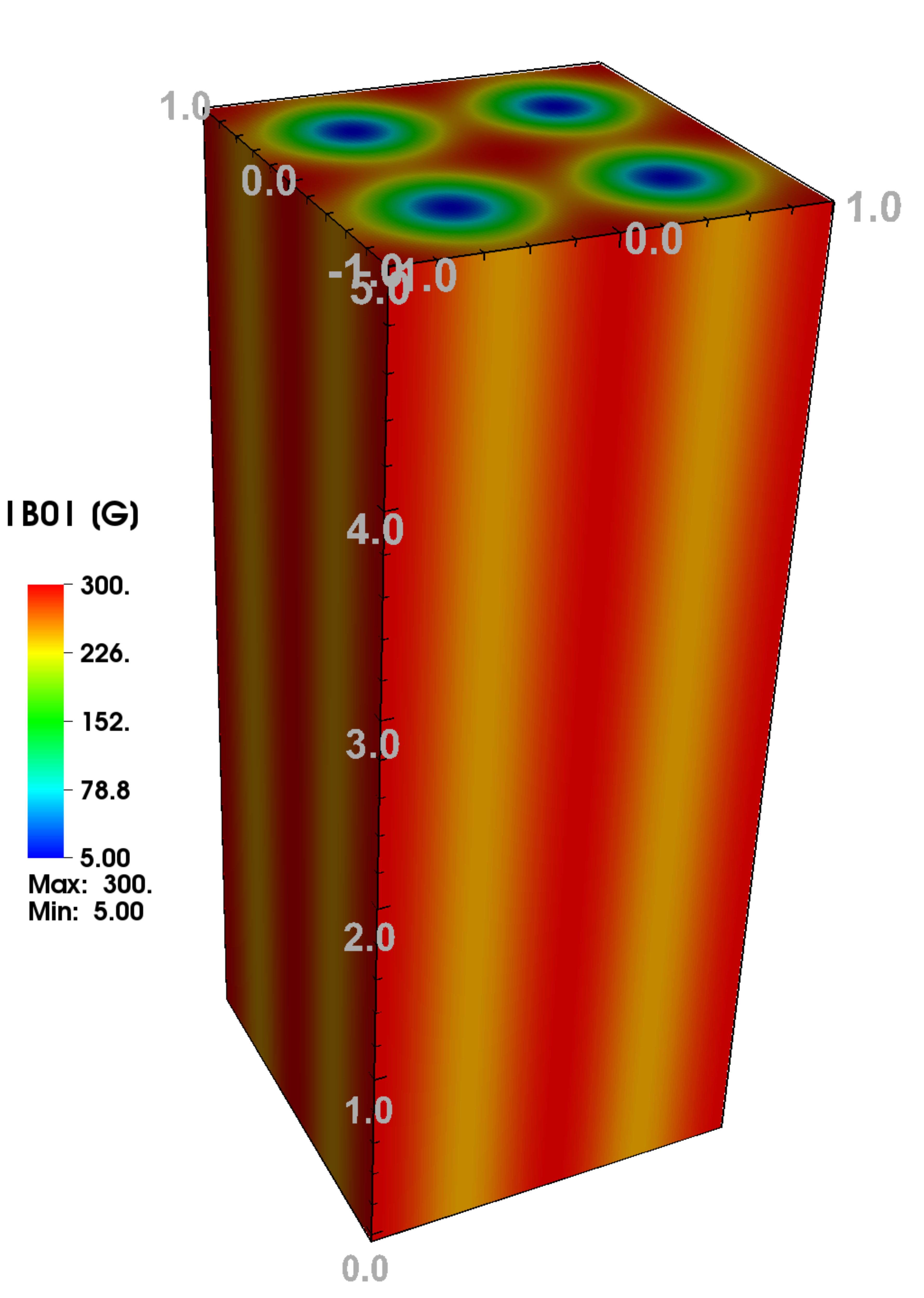} \includegraphics[width=0.33\textwidth]{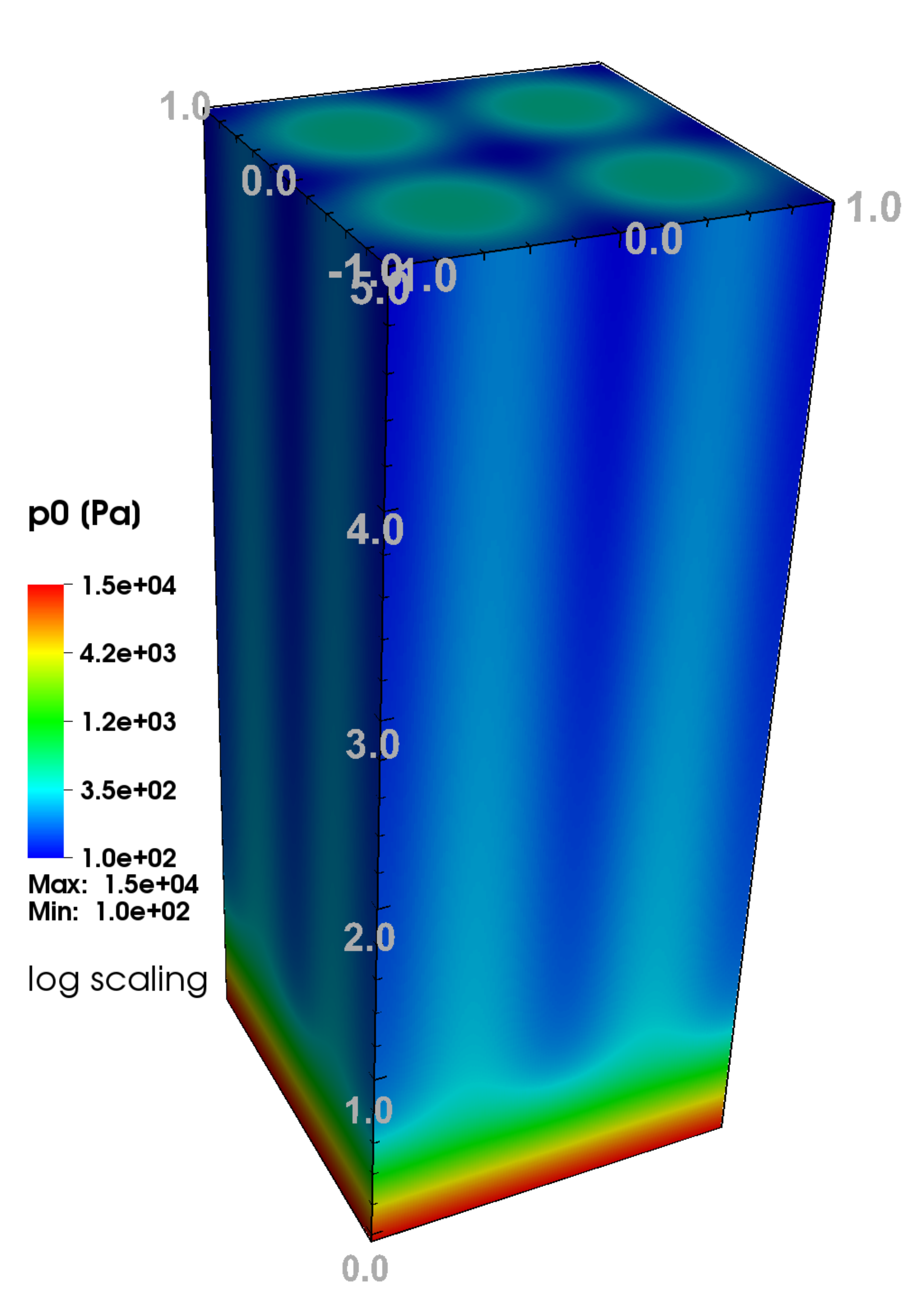} \includegraphics[width=0.33\textwidth]{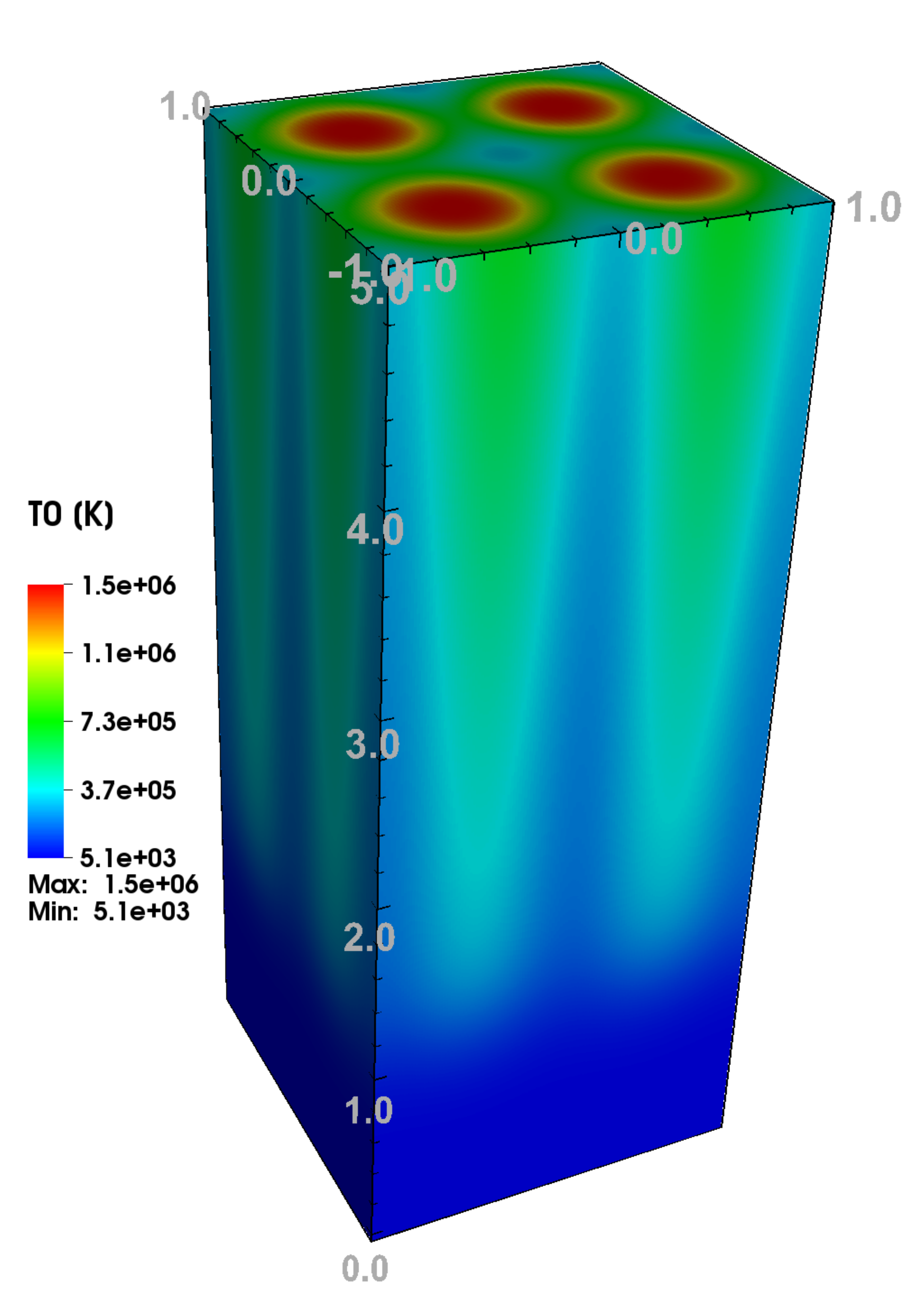}
\includegraphics[width=0.33\textwidth]{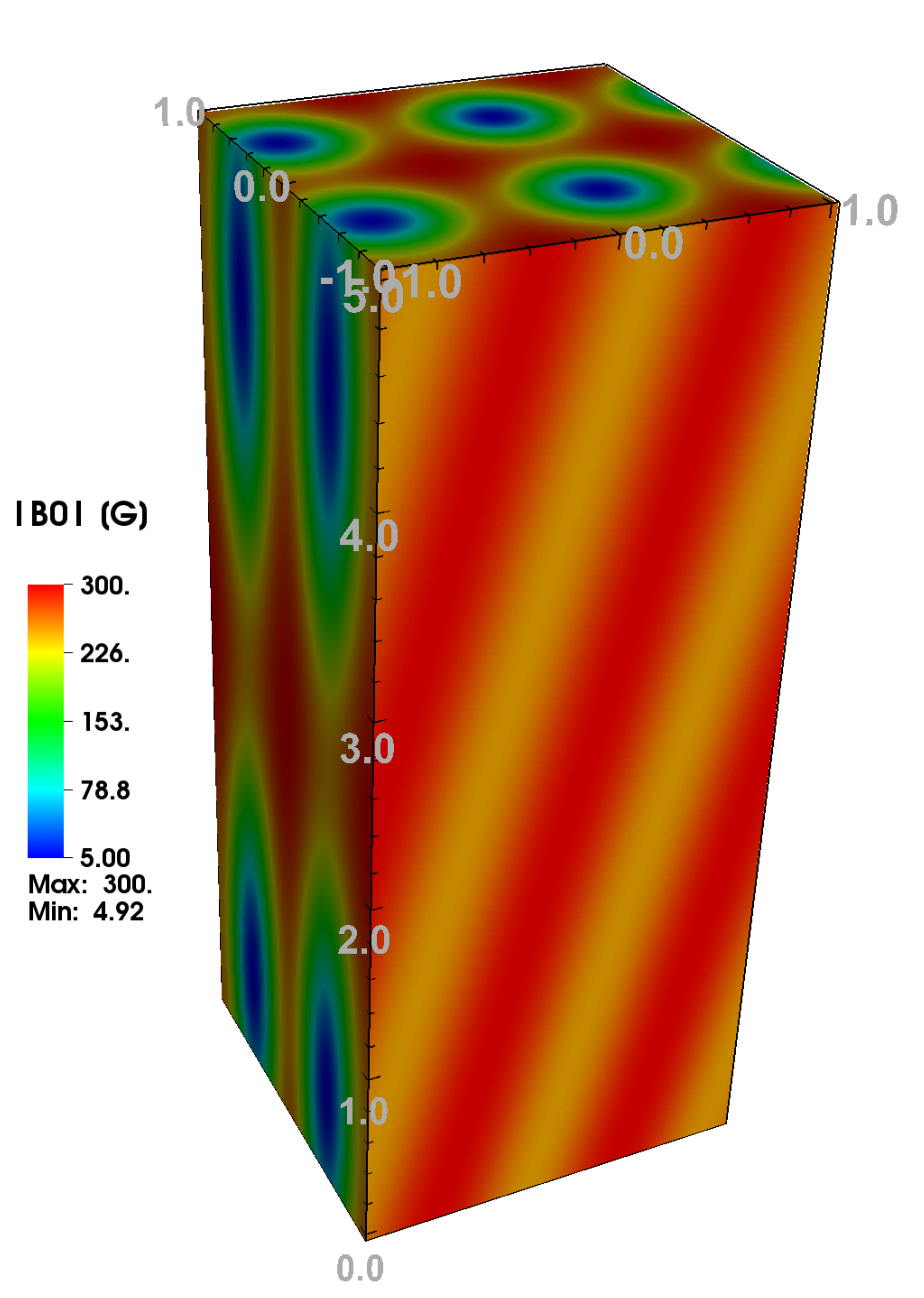} \includegraphics[width=0.33\textwidth]{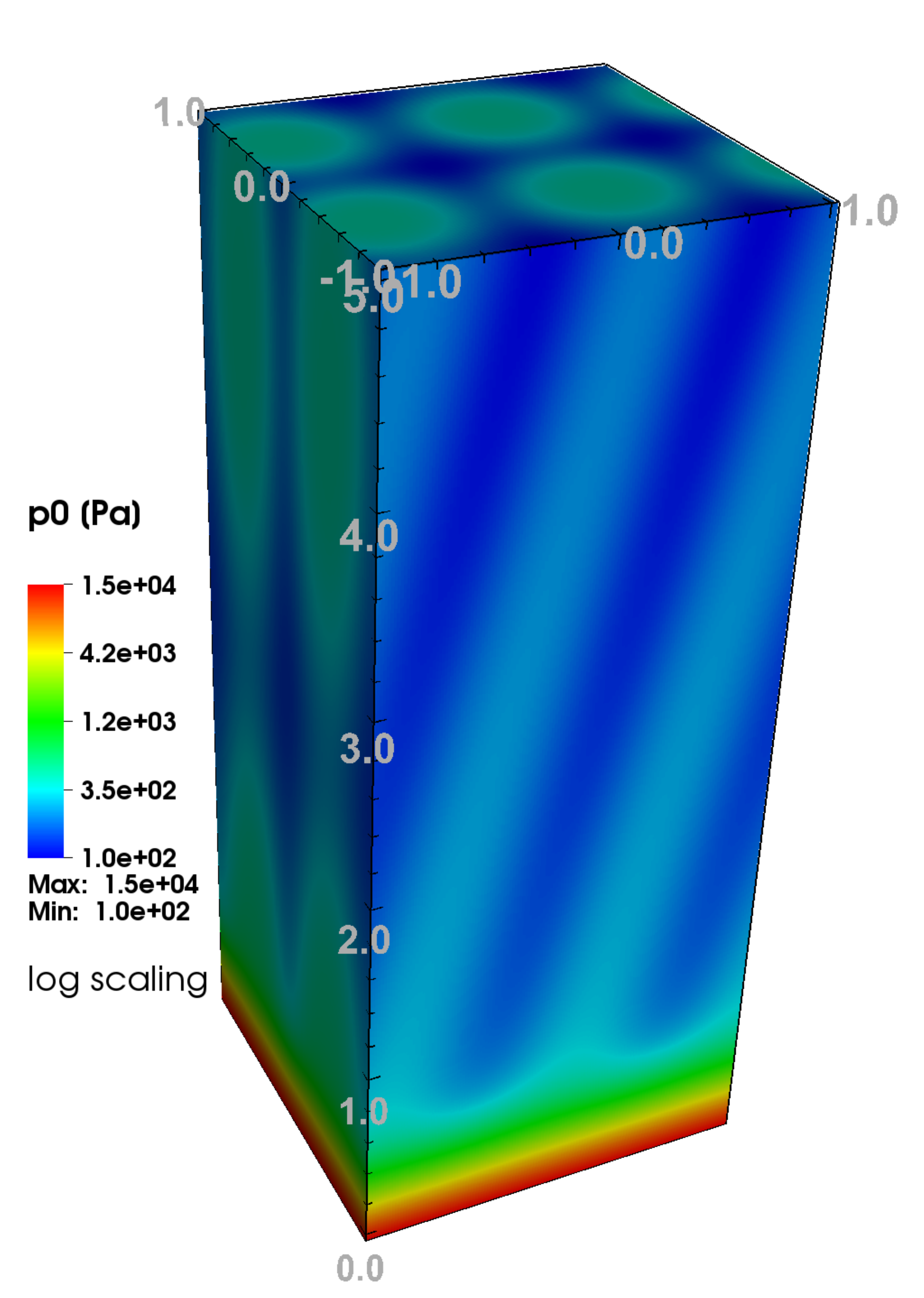} \includegraphics[width=0.33\textwidth]{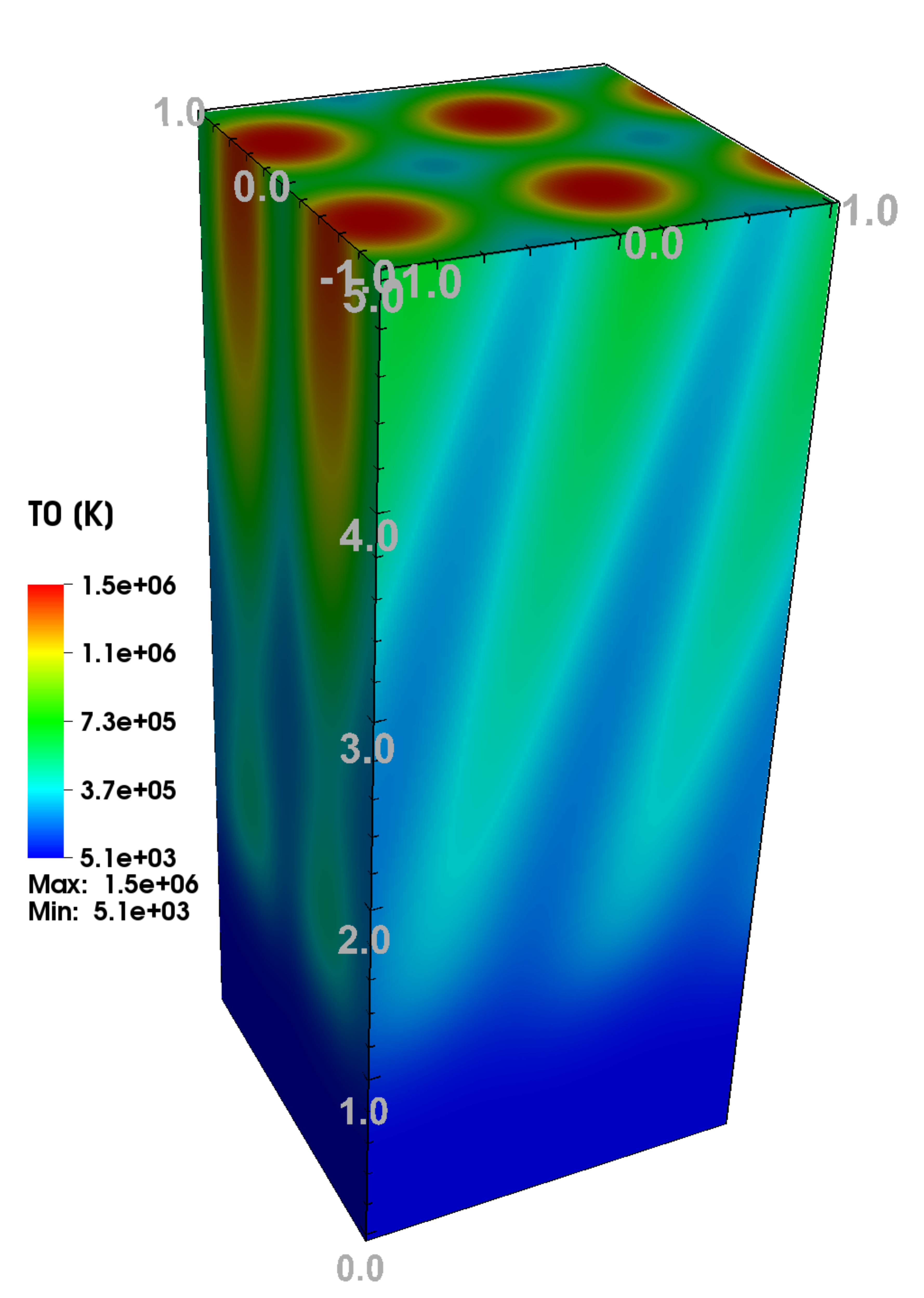}
\caption{Magnetohydrostatic model of four loops for $\theta=0^{\circ}$ (top row) and $\theta=15^{\circ}$ (bottom row).  \textit{Left column:} Magnitude of the total magnetic field. \textit{Middle column:} Pressure (logarithmic scaling). \textit{Right column:} Temperature. All length scales are in Mm. \label{fig:atmosphere}}
\end{figure*}

In the first step we define a divergence-free magnetic field, where the total field strength forms several straight loops of reduced magnetic field with a gauss-shaped cross section
\begin{equation} \label{eq:bxby_vertical}
  B_{0,x}=B_{0,y}=0,
\end{equation}
\begin{equation} \label{eq:bz_vertical}
  B_{0,z}=a \left[ 1-\sum_i^n{\exp \left\lbrace - \frac{ \left( x-x_{0,i} \right) ^2+ \left( y-y_{0,i} \right) ^2}{\sigma_i^2} \right\rbrace } \right] +b.
\end{equation}
The sum corresponds to a sum over all loops, where $x_{0,i}$ and $y_{0,i}$ describe the coordinates of the loop centers and $\sigma_i$ define the thickness of each loop. We place four loops evenly inside the domain with a distance of 1 Mm in the $x$- and the $y$-direction between the loops and use $\sigma=1/3$ Mm for all $i$. The constants $a=309$ G and $b=5$ G define the strength of the magnetic field. The magnetic field has its minimum in the loop centers with value $b$, whereas the theoretical maximum outside the loops is $a+b$. However, owing to the tight structuring of our loops, this maximum is never reached. The resulting magnetic field ranges from 5 G to 300 G and is shown in Fig. \ref{fig:atmosphere} at the top left.  Since we only have a magnetic field component in the $z$-direction and the magnetic field does not change with height, $\vec{\nabla} \cdot \vec{B_0}=0$ is automatically fulfilled.

To study the effect of an inclined magnetic field as well, we rotate the vertical loop system from Equations \ref{eq:bxby_vertical} and \ref{eq:bz_vertical} clockwise around the y-axis by an angle $\theta$, which modifies the equations to
\begin{eqnarray}
  B_{0,x}&=&\tilde{B} \sin(\theta), \\
  B_{0,y}&=&0, \\
  B_{0,z}&=&\tilde{B} \cos(\theta),
\end{eqnarray} 
with
\begin{equation}
  \tilde{B}= a \left[ 1- \sum_i^n \exp \left\lbrace - \phi_i \right\rbrace \right] +b,
\end{equation}
and
\begin{equation}
  \phi_i= \frac{ \left( -\sin \left( \theta \right) z+\cos \left( \theta \right) x-x_{0,i} \right)^2 }{\sigma_i^2}+ \frac{ \left( y-y_{0,i} \right)^2}{\sigma_i^2}.
\end{equation}
We note that  to keep the domain periodic, we extend the it in the $x$-direction and therefore also slightly change the corresponding resolution. That leads to a number of points in the $x$-direction of $n_x=$145 for $\theta = 15^{\circ}$. The total magnetic field for the case of $\theta = 15^{\circ}$ is shown in Fig. \ref{fig:atmosphere} at the bottom left.

\begin{figure*}[]
\centering
\includegraphics[scale=0.5]{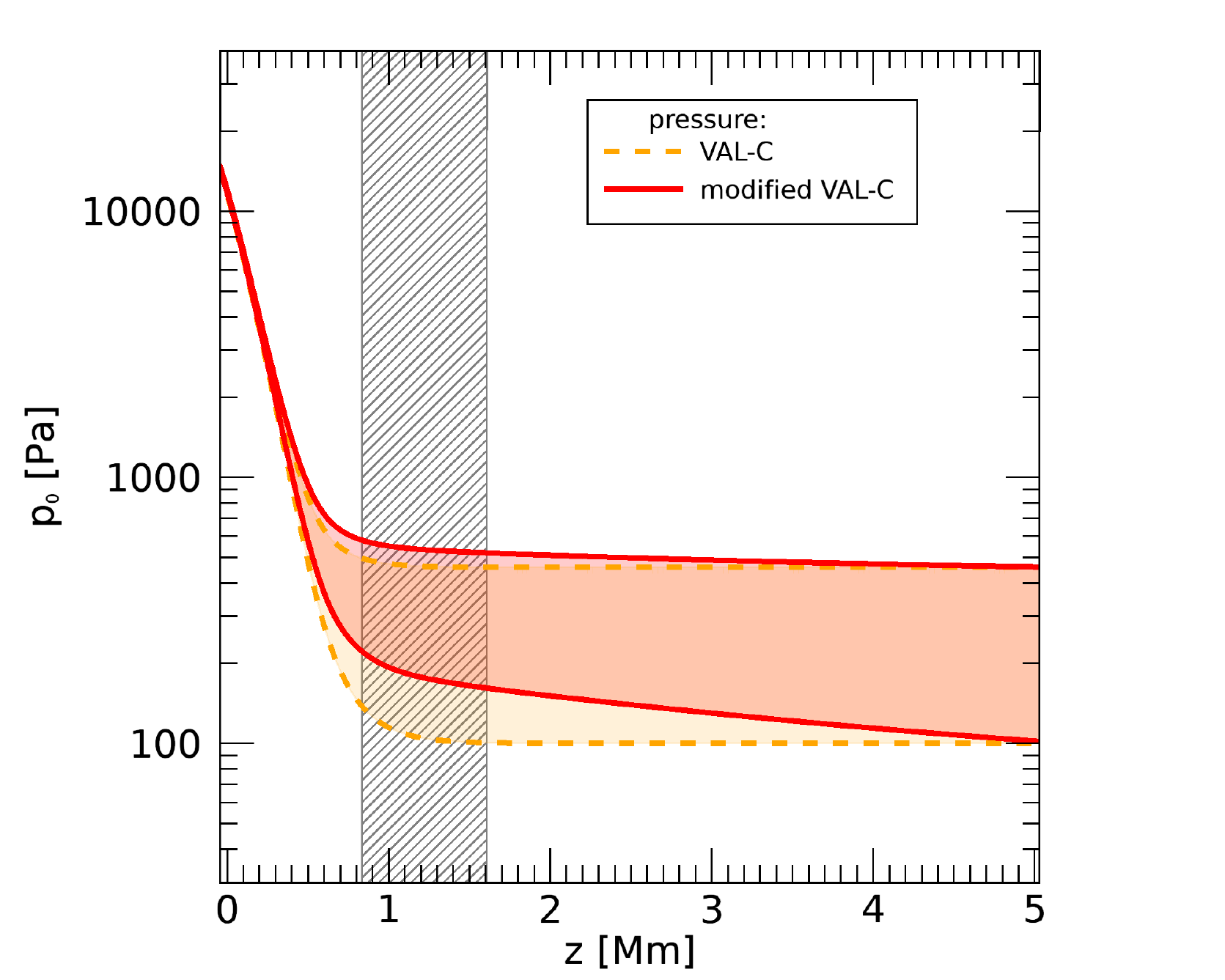}\includegraphics[scale=0.5]{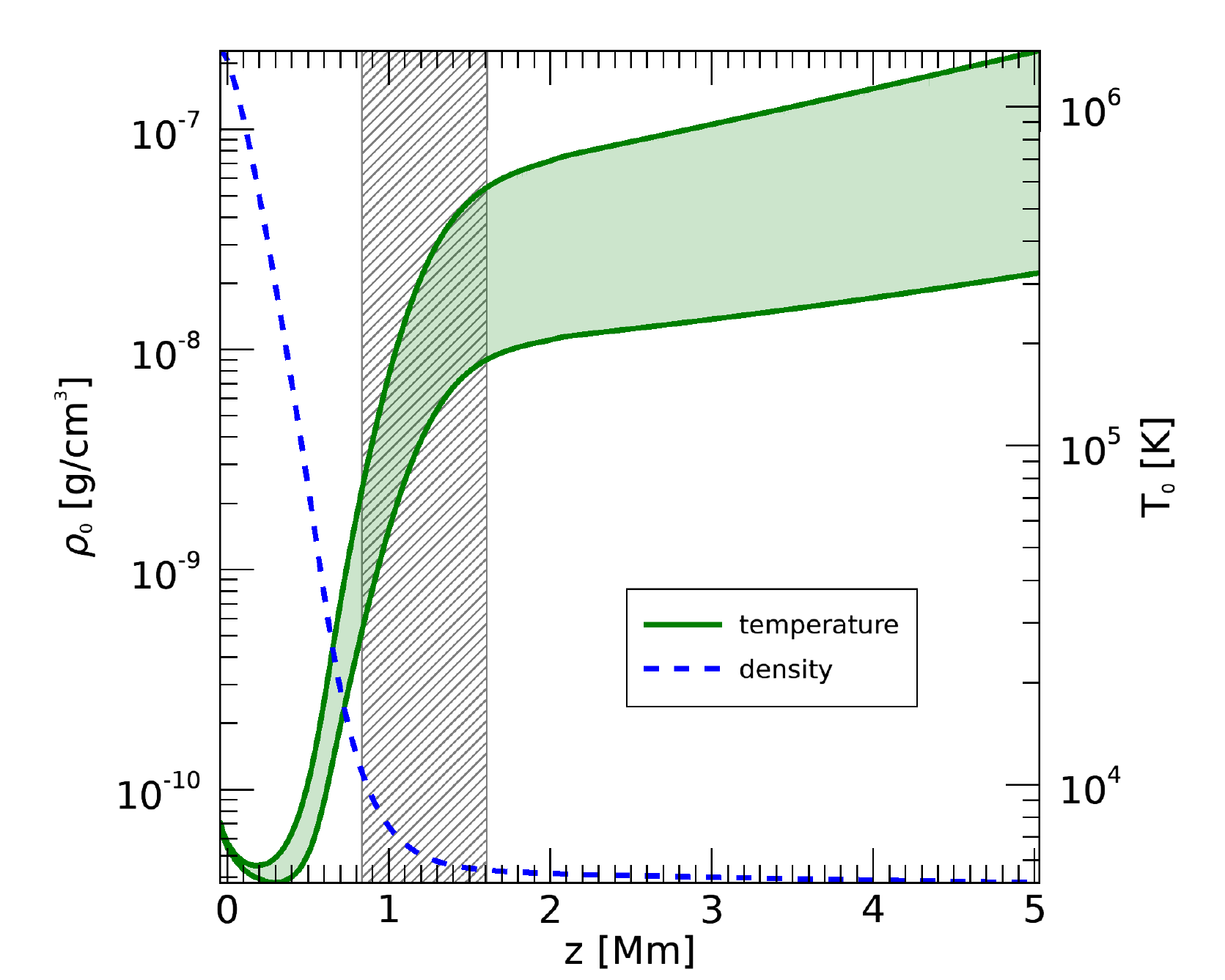}
\caption{Profiles and profile ranges of the atmospheric quanities for the vertical case as a function of $z$. The lines indicate the minimum and maximum values at each height, respectively. The striped region indicates the transition region. \textit{Left:} Pressure according to the VAL-C model (orange dashed lines) and with added exponential term (red solid lines). \textit{Right:} Temperature (green solid lines) and density (blue dashed line, constant in $x$ and $y$). \label{fig:profiles}}
\end{figure*}

In order to calculate the pressure and density we split Equation \ref{eq:mhs_equilibrium} into its three components and slightly reorder these components as follows:
\begin{multline}\label{eq:I}
\mathbf{I}:  B_{0,y} \partial_y B_{0,x}+B_{0,z} \partial_z B_{0,x} -B_{0,y} \partial_x B_{0,y}-B_{0,z} \partial_x B_{0,z} \\ = \mu_0 \partial _x p_0, 
\end{multline}
\begin{multline}\label{eq:II}
\mathbf{II}:  B_{0,x} \partial_x B_{0,y}+B_{0,z} \partial_z B_{0,y} -B_{0,x} \partial_y B_{0,x}-B_{0,z} \partial_y B_{0,z} \\ = \mu_0 \partial _y p_0, 
\end{multline}
\begin{multline}\label{eq:III}
\mathbf{III}:  B_{0,x} \partial_x B_{0,z}+B_{0,y} \partial_y B_{0,z} -B_{0,x} \partial_z B_{0,x}-B_{0,y} \partial_z B_{0,y} \\ = \mu_0 \partial _z p_0 + \rho_0 g \mu_0, 
\end{multline}
where $\partial x_j=\partial/\partial x_j, j={1,2,3}$. Reforming and integrating the first two components gives us
\begin{equation} \label{eq:pressure_calc}
  p_0=\tilde{p}_I+f_1(y,z)=\tilde{p}_{II}+f_2(x,z)=\tilde{p}_I+h(z),
\end{equation}
where $\tilde{p}_I$ and $\tilde{p}_{II}$ are the pressures calculated from integrating Equation \ref{eq:I} and Equation \ref{eq:II}, respectively, and $f_1$ and $f_2$ are functions resulting from the integral that have to be determined. For our magnetic field model $\tilde{p}_I=\tilde{p}_{II}$, which leads to the last equality of Equation \ref{eq:pressure_calc}. The resulting expression for the pressure is
\begin{equation} \label{eq:pressure_expression}
  p_0=- \frac{1}{2 \mu_0} \tilde{B}^2+h(z).
\end{equation}
The function $h(z)$ is arbitrary, as it has no influence on Equations \ref{eq:I} and \ref{eq:II}, and represents the vertical pressure stratification, and the constant that has to be added to make Equation \ref{eq:pressure_expression} positive. However, we stress the significance of choosing $h(z)$ wisely, as this term essentially determines our vertical density profile. Therefore, we define the vertical pressure stratification according to the VAL-C model and add an exponential term to modify the stratification
\begin{multline} \label{eq:pressure_strat}
  h(z)=p_{\mathrm{bot}} \exp \left\lbrace - \int \frac{1}{H(z)} dz \right\rbrace+ \\ 150 \mathrm{[Pa]} \exp \left\lbrace - \frac{z}{600 \cdot 6046 \mathrm{[m]}} +0.015 \right\rbrace + 422\mathrm{[Pa]}.
\end{multline}
In this case, $p_{\mathrm{bot}}=14$ kPa is the pressure at the bottom of the domain for the first term of Equation \ref{eq:pressure_strat}. The scale height $H(z)$ is calculated with a temperature profile that follows the VAL-C model until the transition region and has a constant temperature for the corona. The two regions of the temperature profile are connected by a cosine-shaped transition region. Since this initial temperature profile is only used for calculating the vertical pressure stratification and is changed in the next step, we abstain from mentioning the exact expression. The constants in the exponential term of Equation \ref{eq:pressure_strat} are related to our practical implementation and their exact values have no deeper meaning. Figure \ref{fig:profiles} shows the pressure profile as a function of height for $\theta=0\degr$ for pressure according to the VAL-C model (dashed lines) and for pressure with added exponential term, as used in our model (solid lines). Both profiles have the same start and end points, but in the modified version the slope in the upper part of our model is larger, which leads to higher density values in that region. This is necessary as a consequence of the constant magnetic field along the loops the density would be very low otherwise, which would lead to unreasonably high temperatures. However, the additional term also leads to a strong broadening of our transition region, which we deem a compromise that has to be made.

As soon as the magnetic field and the pressure are known it is straight forward to get an expression for $\rho$ from Equation \ref{eq:III}. Finally, we calculate the temperature from the ideal gas law. Figure \ref{fig:profiles} (right) shows the density (dashed) and temperature (solid) profiles as a function of $z$ for the vertical case, whereas Fig. \ref{fig:atmosphere} shows a 3D plot of the pressure and temperature for both the vertical and the $15^{\circ}$ inclined case. We note that the pressure is vertically much more stratified than horizontally structured.

\begin{figure*}
\centering
\includegraphics[width=0.33\textwidth]{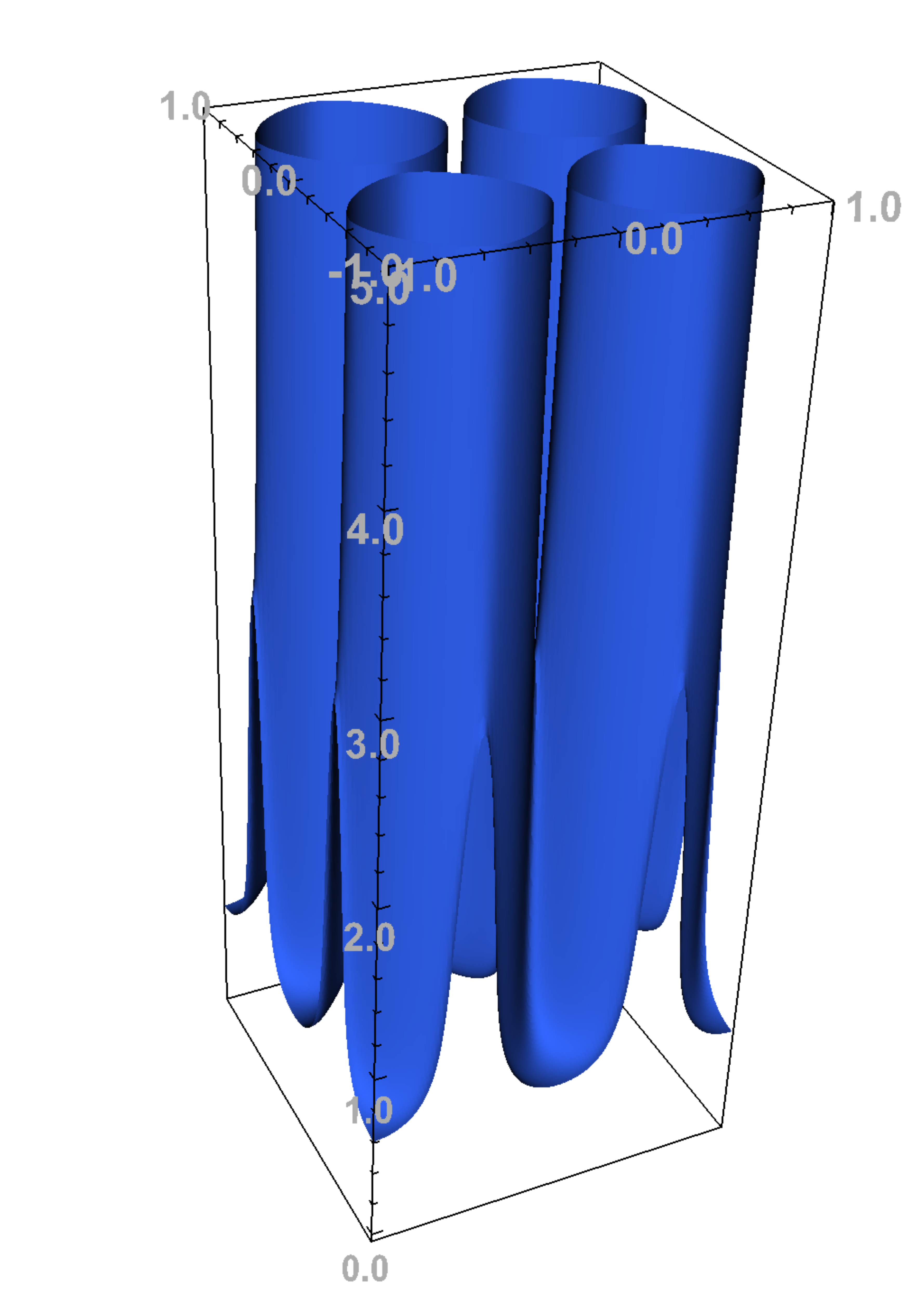} \includegraphics[width=0.33\textwidth]{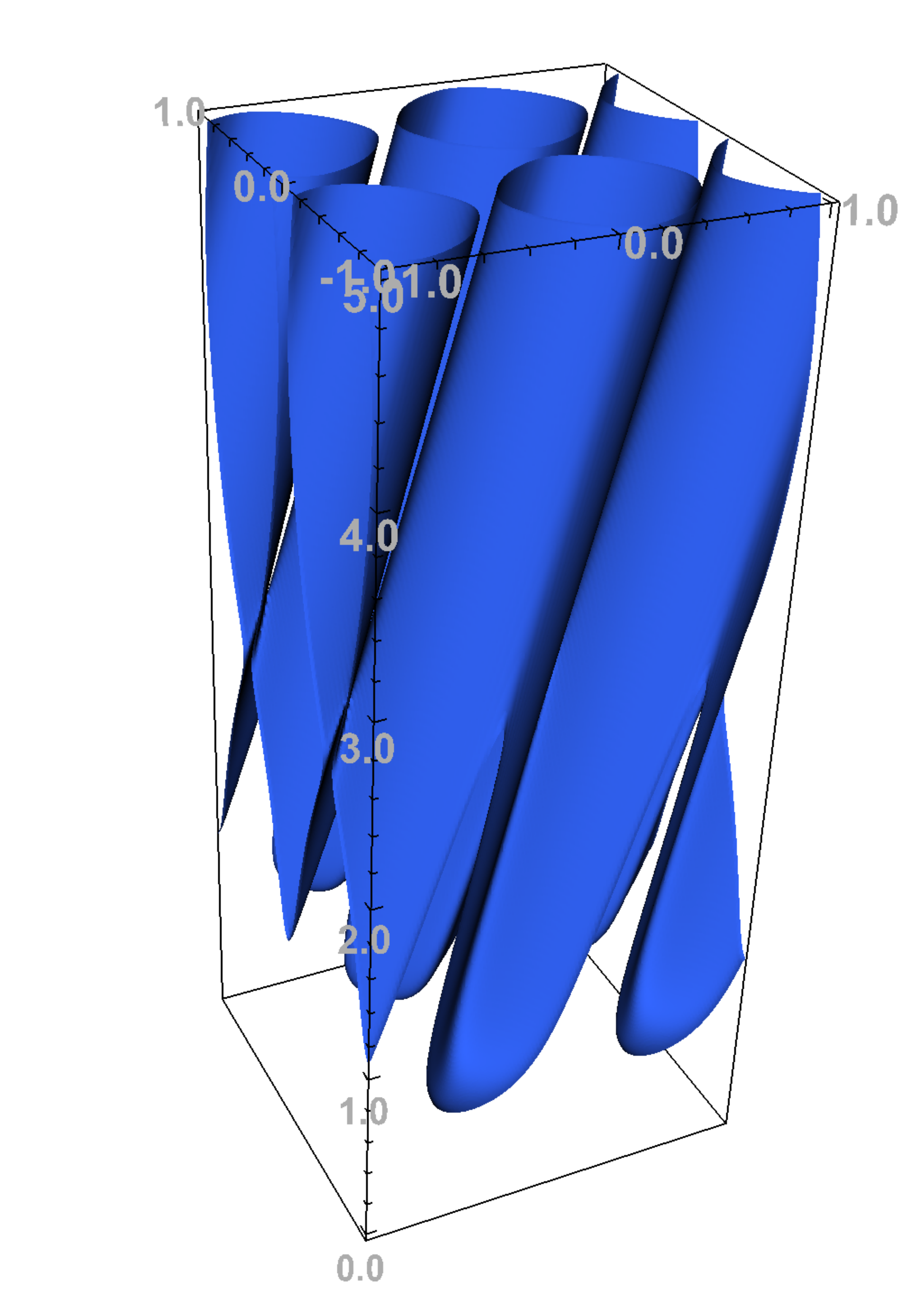}
\caption{Contour of $\beta=1$ for $\theta=0^{\circ}$ (left) and $\theta=15^{\circ}$ (right). At the bottom of the domain (photosphere) $\beta \gg 1$. \label{fig:beta}}
\end{figure*}

Since the pressure is higher in the loop interior, where the magnetic field is lower, we get a sound speed profile that is much higher inside the loops than outside. On the contrary, the Alfv\'{e}n speed has its maximum outside the loops, while being very low in their center. This would be expected for a coronal loop with higher density than its surroundings, however, to fulfill Equation \ref{eq:mhs_equilibrium} the density is horizontally constant in our model. This is easily visible in Equation \ref{eq:III} considering the vertical case with $B_{0,x}=B_{0,y}=0$ and taking into account that $\partial_z p_0$ is independent of $x$ and $y$ for $\theta=0\degr$. The pressure and magnetic field distributions lead to a horizontally strongly structured plasma-$\beta$ profile. The $\beta=1$ contour is plotted in Fig. \ref{fig:beta}. As we go up in the atmosphere, the plasma-$\beta$ decreases in the loop exterior already below the transition region to lower than unity, while it is always much higher than unity in the loop interior until the top of the domain.

\section{Numerical setup} \label{sec:numerics}

For our simulations we used the MANCHA3D code \citep{khomenko_collados_2006,felipe_etal_2010,khomenko_etal_2018} developed at the Instituto de Astrof\'{i}sica de Canarias in Tenerife, Spain. This code solves the fully nonlinear magnetohydrodynamics equations for perturbed variables, which is why we initially had to define a MHS equilibrium to perturb it in the simulations. For the present work we consider an adiabatic system and neglect changes of the mean molecular weight due to ionization.

On the vertical faces of the computational box we set periodic boundary conditions. Our system can therefore be viewed as an ensemble of thin loops (or loop strands) extending infinitely to all horizontal sides that reasonably represent groups of spicules. To allow waves to escape, we set a Neumann-type zero-gradient open boundary condition at the upper boundary. At the first seven layers of the lower boundary we applied a driver that follows an analytic solution for a vertical gravity-acoustic wave \citep{mihalas_mihalas_1984}. The driver is described in detail by \cite{santamaria_etal_2015} and in this work we only repeat  the general form,
\begin{equation}
  v_{z,1}=V_0 \exp \big \lbrace \frac{z}{2H}+k_{zi}z \big \rbrace \sin(\omega t -k_{zr}z),
\end{equation}
\begin{equation}
  \frac{p_1}{p_0}=V_0 |P| \exp \big \lbrace \frac{z}{2H}+k_{zi}z \big \rbrace \sin(\omega t -k_{zr}z +\phi_P) ,
\end{equation}
\begin{equation}
  \frac{\rho_1}{\rho_0}=V_0 |R| \exp \big \lbrace \frac{z}{2H}+k_{zi}z \big \rbrace \sin(\omega t -k_{zr}z +\phi_R) ,
\end{equation}
where $v_{z,1}$, $p_1$, and $\rho_1$ are the velocity perturbation in the $z$-direction, pressure perturbation, and density perturbation, respectively. The values $|P|$ and $|R|$ are the relative amplitudes, while $\phi_P$ and $\phi_R$ are the phase-shifts of pressure and density perturbation compared to the velocity perturbation.\ The value $H$ is the pressure scale height and $V_0$ is the amplitude of the velocity perturbation. The vertical wave number $k_z$ is either complex or real, depending on the frequency $\omega$ compared to the isothermal acoustic cutoff frequency $\omega_c$,
\begin{equation}
  k_z=k_{zr}+ik_{zi}=\frac{\sqrt{\omega^2-\omega_c^2}}{c_s}
\end{equation}
with
\begin{equation}
  \omega_c=\frac{\gamma g}{2c_s},
\end{equation}
where $c_s$ is the sound speed and $\gamma=5/3$ is the adiabatic index.

In this work we used a small perturbation of $V_0=10^{-2}$ m/s to stay in the linear regime. In addition, we used a period of 100 s, which leads to a frequency of $\omega\approx0.063$ rad/s. The main reason for choosing this small period is that the imperfect open boundary conditions at the top cause waves to reflect and propagate downward. The chosen period allows us to study at least half of a wave period before the reflected waves interfere. For this period, $\omega > \omega_c$ is valid for the whole domain, so the waves excited from the bottom of the domain never reach a cutoff region. To put it into a solar context, the p-modes in our simulations are no longer trapped within the solar interior and can propagate through the chromosphere. Since the waves are not trapped within a resonant cavity, their amplitude is not amplified by constructive interference, which validates our choice of a small $V_0$. We plan to use larger periods that allow a cutoff region in future work.

\section{Methods for data interpretation} \label{sec:methods}

\subsection{Decomposition into components} \label{subsec:decomposition}

In order to distinguish the different wave modes, it is necessary to bring our simulation data into a form that allows us to visualize the characteristics of the expected modes. \cite{tarr_etal_2017} decomposed their data into kinetic, acoustic, and magnetic energy densities, which allowed these authors to decouple fast from slow waves and magnetic from acoustic waves. Another decomposition method was carried out by \cite{khomenko_etal_2018}, who, following \cite{cally_2017}, constructed three quantities based on the physical properties of the waves: (1) $f_\mathrm{alf}$ for the incompressible perturbation propagating along the magnetic field, (2) $f_\mathrm{long}$ for the compressible perturbation propagating along the magnetic, and (3) $f_\mathrm{fast}$ for the compressible perturbation perpendicular to the magnetic field. While $f_\mathrm{alf}$ is a quantity describing the Alfv\'{e}n waves for all $\beta$, $f_\mathrm{long}$ and $f_\mathrm{fast}$ decouple the slow and the fast magneto-acoustic waves only for $\beta<1$.

However, since we have both $\beta<1$ and $\beta>1$ regions in our model and expect tube modes to be excited, we adopted the approach of \cite{mumford_etal_2015}, who split the velocities and fluxes into three orthogonal components defined by the magnetic flux surfaces. These components are defined by the unit vectors longitudinal ($\vec{\hat{e}}_\parallel$), azimuthal ($\vec{\hat{e}}_a$), and normal ($\vec{\hat{e}}_\perp$) to those surfaces. Since we only used small perturbations, magnetic flux surfaces and therefore also the resulting unit vectors for the three components are constant in time. The unit vector for the longitudinal component $\vec{\hat{e}}_\parallel$ points into the direction of the magnetic field and is therefore easy to compute. Because all field lines in our model are straight, this vector is the same for all points of the domain.

\begin{figure}[]
\centering
\includegraphics[width=0.5\textwidth]{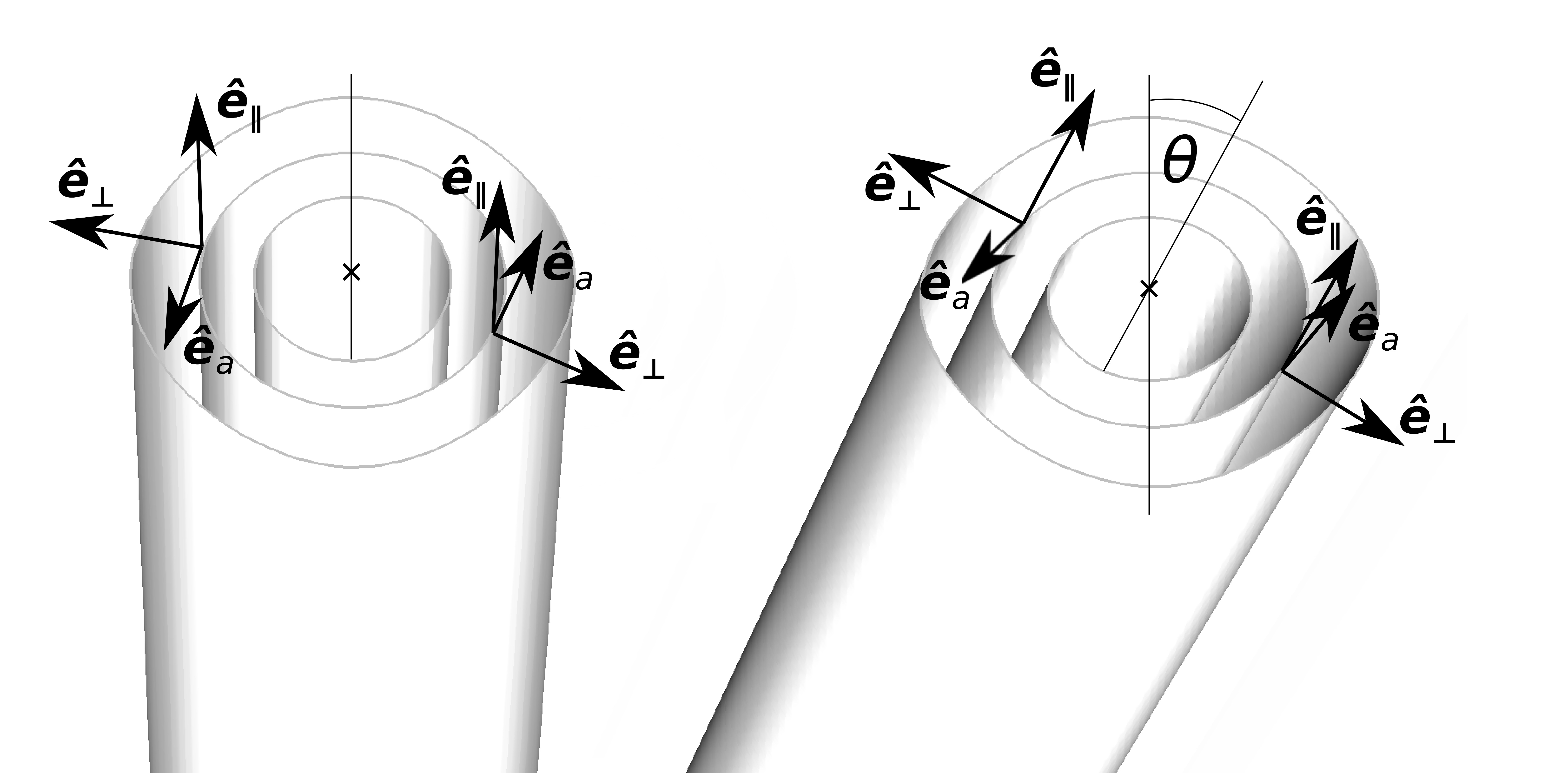}
\caption{Orthogonal decomposition vectors parallel ($\vec{\hat{e}}_\parallel$), normal ($\vec{\hat{e}}_\perp$), and azimuthal ($\vec{\hat{e}}_a$) to the magnetic iso-surfaces (cylindrical shapes) for a vertical (left) and inclined (right) loop. \label{fig:pna_cartoon}}
\end{figure}

Calculating the azimuthal unit vector $\vec{\hat{e}}_a$ proves to be more difficult. We solve it by calculating the 2D isocontour of the magnetic field for all horizontal layers and fitting a straight line to the contour for each pixel. The direction of the linear fit is then the direction of $\vec{\hat{e}}_a$. For the inclined case $\vec{\hat{e}}_a$ gets an appropriate vertical component to be normal to $\vec{\hat{e}}_\parallel$. However, with this method all azimuthal vectors point to the positive $x$-direction, which results in half of the vectors being clockwise (top half of the loops in respect to $y$), while the others are counterclockwise (bottom half of the loops in respect to $y$). Because of the regular structure of the loop system it is simple to distinguish those regions. We multiply all clockwise azimuthal unit vectors with -1 to have a consistent sense of direction. Finally, the normal unit vector $\vec{\hat{e}}_\perp$ is calculated by $\vec{\hat{e}}_\perp=\vec{\hat{e}}_a \times \vec{\hat{e}}_\parallel$. With this convention, $\vec{\hat{e}}_\parallel$ always points upward (and for the inclined case also into the positive $x$-direction), $\vec{\hat{e}}_a$ is parallel to the flux surfaces and points to the counterclockwise direction, and $\vec{\hat{e}}_\perp$ points away from the loop centers. Figure \ref{fig:pna_cartoon} sketches the vector directions for vertical and inclined loops.

\subsection{Expected wave modes} \label{subsec:expected_modes}

\begin{figure}
\centering
\includegraphics[scale=0.4]{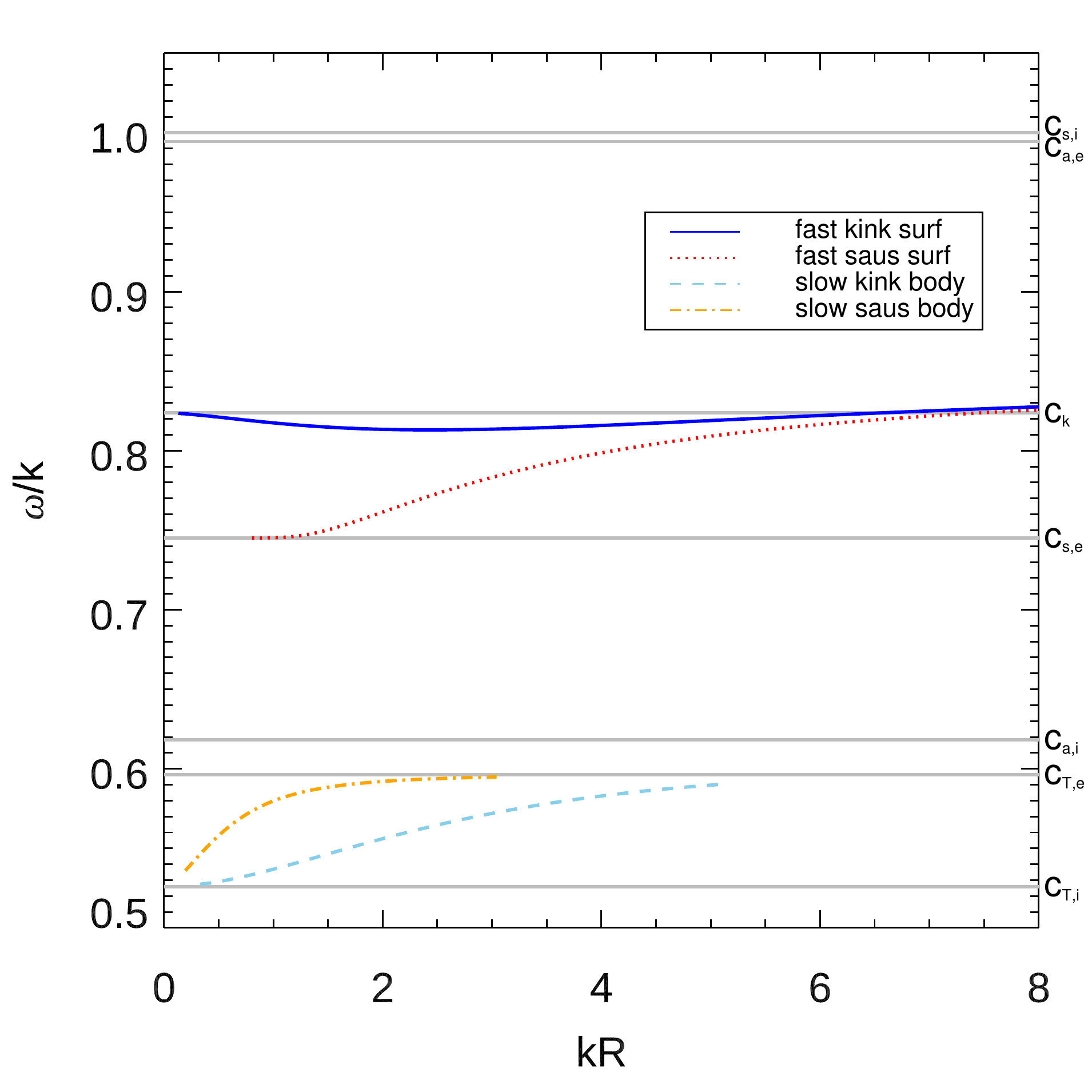}
\caption{Phase speed diagram of sausage modes and kink modes for conditions similar to the model atmosphere at a height of 2 Mm. The horizontal gray lines indicate various characteristic speeds. \label{fig:phase_speed_diagram}}
\end{figure}

Since our atmosphere in not horizontally uniform but has a cylindrical shape, we would also expect the wave modes excited in our simulations to be those of a plasma with cylindrical shape, such as the $m=0$ sausage mode and $m=1$ kink mode, where $m$ is the azimuthal wave number. To approximately calculate what wave modes would appear in our setup, we assume a simple vertically constant cylinder with radius $R$ and internal values $f_i$ embedded in an external plasma with values $f_e$. Similar to our atmosphere, we assume that the magnetic field is parallel to the loop axis. We then followed the mathematical framework of \cite{moreels_vandoorsselaere_2013}. As external and internal values for this simplified model we used the average external and average internal values of our atmosphere at a height of 2 Mm, where the loop boundary is defined by the $\beta=1$ layer. If $c_{s,i}$ is the internal sound speed, this leads to an external sound speed of $c_{s,e}=0.745 c_{s,i}$, an internal Alfv\'{e}n speed of $c_{A,i}=0.618 c_{s,i}$, and an external Alfv\'{e}n speed of $c_{A,e}=0.994 c_{s,i}$. Figure \ref{fig:phase_speed_diagram} shows the resulting phase speed diagram, where the internal and external sound and Alfv\'{e}n speeds are indicated by horizontal gray lines. Also plotted are the internal and external tube speed
\begin{equation}
  c_{T,f}=\frac{c_{s,f}c_{A,f}}{\sqrt{c_{s,f}^2+c_{A,f}^2}}
\end{equation}
and the kink speed
\begin{equation} \label{eq:kink_speed}
  c_k=\sqrt{\frac{\rho_{0,i}c_{A,i}^2+\rho_{0,e}c_{A,e}^2}{\rho_{0,i}+\rho_{0,e}}}.
\end{equation}
For both sausage and kink waves, we only get non-leaky solutions for fast surface waves and slow body waves. Slow surface waves are theoretically possible below $c_{T,i}$, but no solutions are found in this region. Fast body modes could occur above $c_{s,i}$, but they are leaky; we indeed find leaky solutions for a kink body mode there for higher $kR$. The phase speed line of the fast sausage surface mode (red dotted line) actually stops where $kR$ is approximately 0.8 as there are no solutions found for low $kR$, not even for leaky waves.

\begin{figure*}
\centering
\includegraphics[scale=0.4]{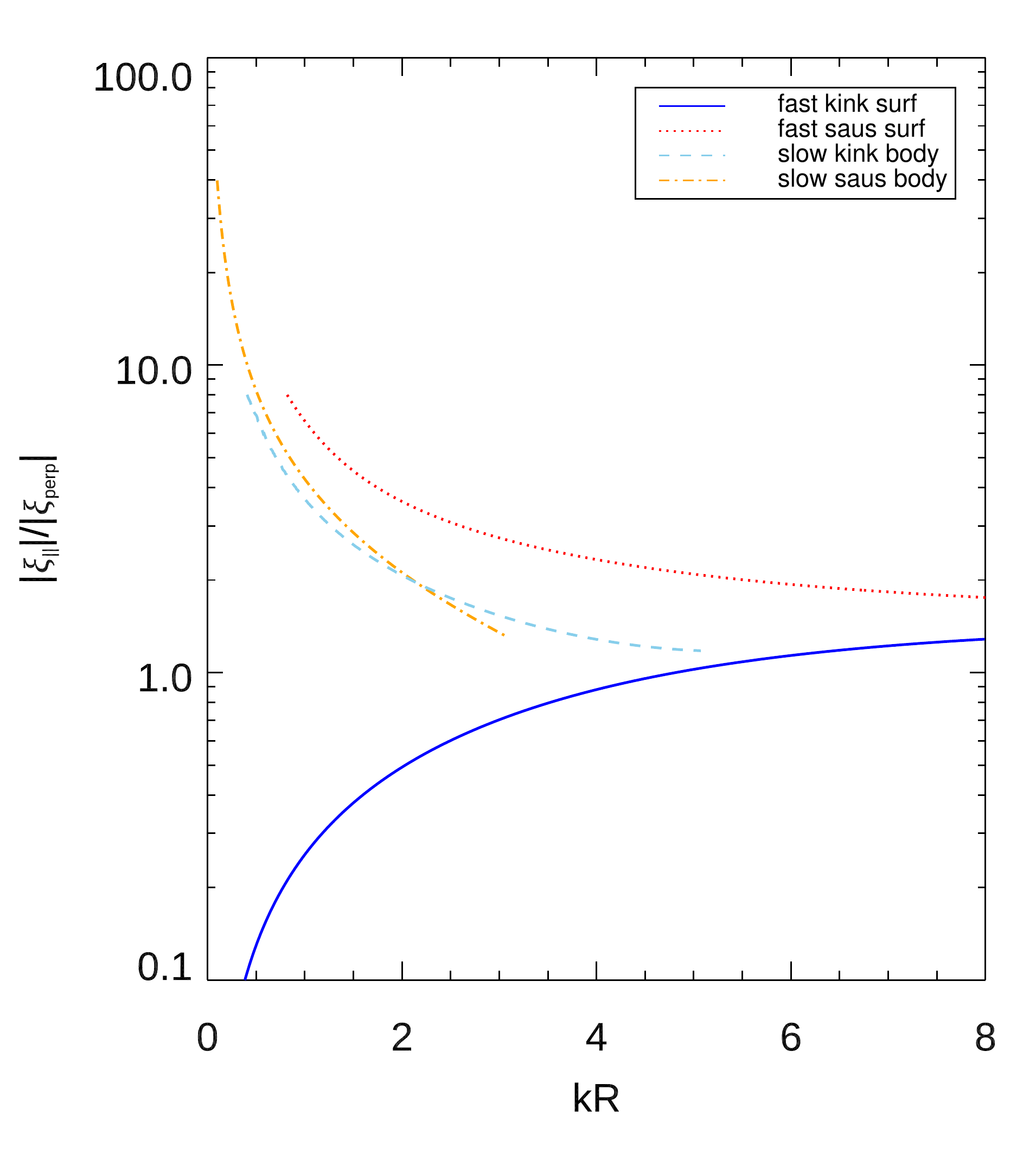}\includegraphics[scale=0.4]{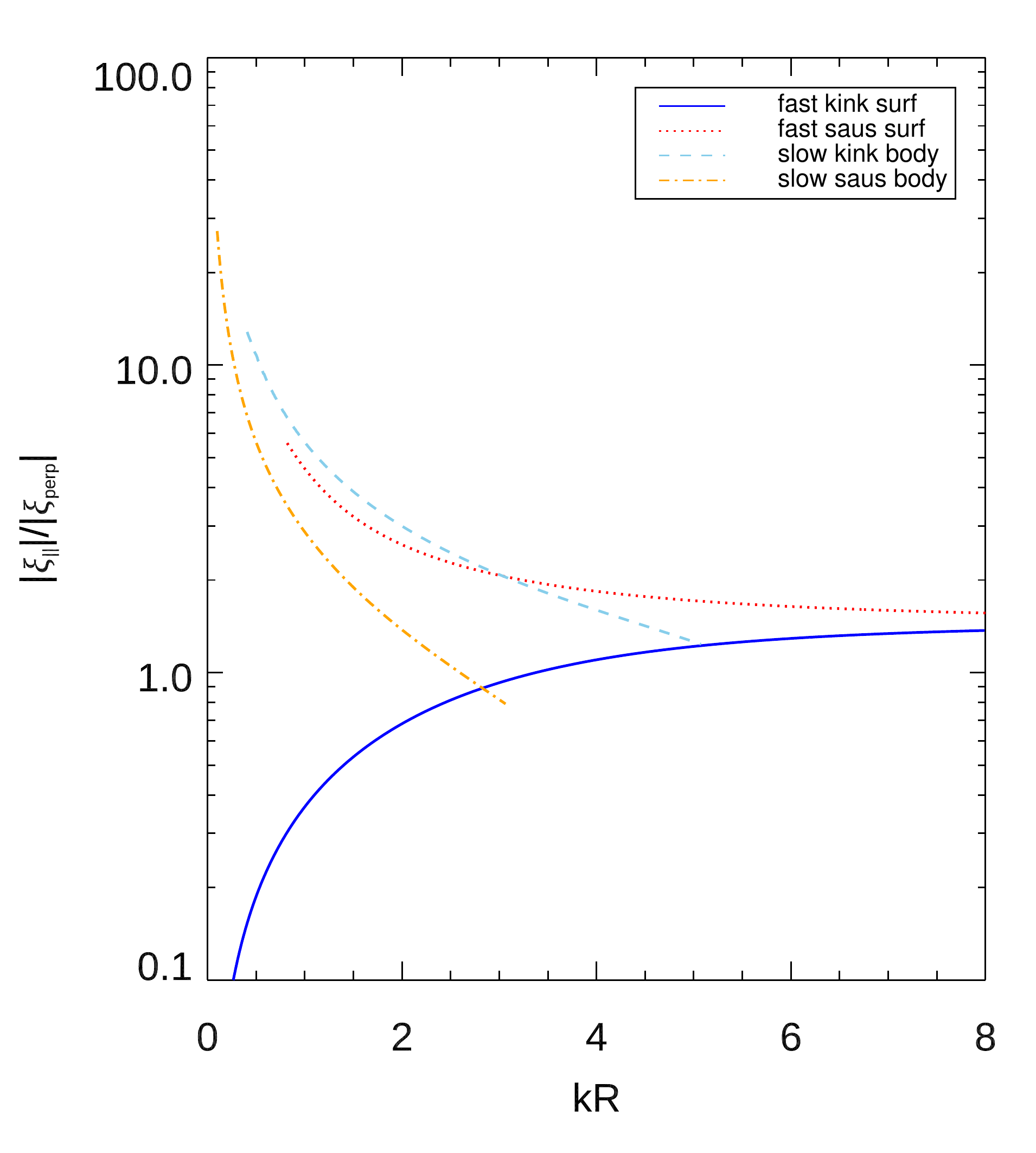}
\caption{Ratio of longitudinal displacement and perpendicular displacement of plasma for sausage and kink modes for conditions similar to the model atmosphere at a height of 2 Mm. \textit{Left:} Inside the loop at $r=0.69R$. \textit{Right:} At the surface of the loop ($r=R$).\label{fig:displacement_ratio}}
\end{figure*}

It is now possible to calculate the magnitude of the ratio of longitudinal displacements to perpendicular displacements for the modes shown in Fig. \ref{fig:phase_speed_diagram}. Unlike the normal component in the decomposition we described in Sect. \ref{subsec:decomposition}, perpendicular displacement describes plasma displacement in all directions perpendicular to the loop axis, so $\xi_{\mathrm{perp}}=\sqrt{\xi_a^2+\xi_\perp^2}$. This ratio depends on the distance to the loop axis $r$ and is shown in Fig. \ref{fig:displacement_ratio} at $r=0.69R$ (left) and at $r=R$ (right). From the figures it is clear that, although the curves slightly change, the general behavior of the displacement ratios stays the same regardless of the chosen point inside the loop or the loop surface. While the slow body modes and the fast sausage surface mode have much higher parallel displacements than perpendicular displacements for small $kR$ compared to larger $kR$, it is the opposite for the fast kink surface mode. This makes the fast sausage surface mode and the fast kink surface mode easily distinguishable from each other.

\section{Results and discussion} \label{sec:results}

\subsection{Wave propagation for vertical flux tubes}

\begin{figure*}
\includegraphics[width=\textwidth]{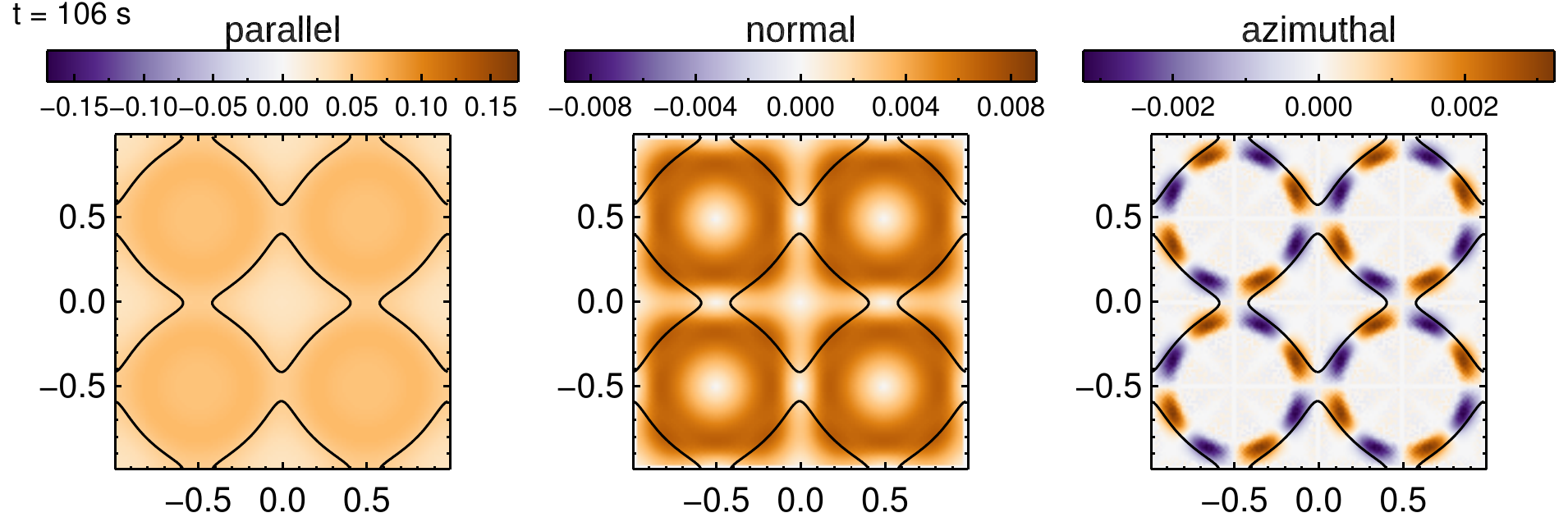}
\caption{Components of the velocity perturbation in a horizontal cut at 2 Mm for $\theta=0^{\circ}$ at a time of 106 s after the start of the simulation. The velocities are given in m/s and the spatial scales are in Mm. The black lines show the $\beta=1$ contour, with $\beta\gg1$ inside and $\beta<1$ outside the loop. The values of the first two pixels next to the margin are set to zero for the normal and azimuthal component, as the corresponding vectors were badly defined in that region. The temporal evolution is available as an online movie. \label{fig:pna_horizontal_0deg}}
\end{figure*}

We first study simulations with the vertical ($\theta=0^{\circ}$) case. Our goal is to investigate the conversion of p-modes that arrive at the corona. For that purpose we look at the evolution of a horizontal cut through the domain at a height of 2 Mm. We later show that we have some issues with wave reflection from the upper boundary,
so investigating the wave behavior at 2 Mm instead of higher up allows us to analyze a longer time sequence before the reflections from the upper boundary intervene. This is possible because the wave behavior does not change much above the transition region after 2 Mm, which is the case even though the high $\beta$ regions inside the loops are still merged together at that height. 
The movie showing the time development of the horizontal cut before the reflections from the upper boundary arrive ($ t \le 170$ s) is available online. A screenshot of the movie is shown in Fig. \ref{fig:pna_horizontal_0deg}. It shows the three velocity components at $t=106$ seconds together with the $\beta=1$ contour. For this time series the longitudinal velocity perturbation is always larger inside the loop than outside. The waves arrive at 2 Mm at approximately the same time; the waves inside the loop arrive slightly earlier. The maximum perturbations of the normal component are more than one order of magnitude smaller than the maximum perturbations of the parallel component. 
In the first part of the time series, the normal component has a positive sign everywhere and thus shows an expansion of the whole loop cross section with some normal velocity components outside the loops as well.
In its first maximum at $t=106$ s (as indicated in Fig. \ref{fig:pna_horizontal_0deg}) the expansion of the loop has similarities with a $m=4$ fluting mode. However, this mode would also require some plasma to flow into the loops at the top, bottom, and sides of the loops. At around $t=114$ s the normal component changes its sign and at $t=128$ s it looks the same as in Fig. \ref{fig:pna_horizontal_0deg} but with changed sign (i.e., contraction instead of expansion). We therefore conclude that the wave the p-modes excited is in fact a $m=0$ sausage mode that is deformed by the tight packing of the grid-like positioned loops. The deformed sausage modes are very similar to a superposition of a $m=0$ sausage mode with a $m=4$ fluting mode.

Immediately apparent in the azimuthal component of Fig. \ref{fig:pna_horizontal_0deg} is the ring-like structure of counter-flowing plasma velocities close to the $\beta=1$ layer. However, this ring structure propagates from outside of the loop inward and is only coincidentally at the position of the $\beta=1$ layer for this screenshot. This propagation inward is only apparent \citep{raes_etal_2017}, since it results from a cone-shaped area of high $c_A$ propagating upward. Those waves are probably Alfv\'{e}n waves that are excited by the first impulse of the driver. Other than that, stationary counter-streaming regions appear around the $\beta=1$ layer with the same periodicity as the normal component. 

\begin{figure}
\centering
\includegraphics[scale=0.3]{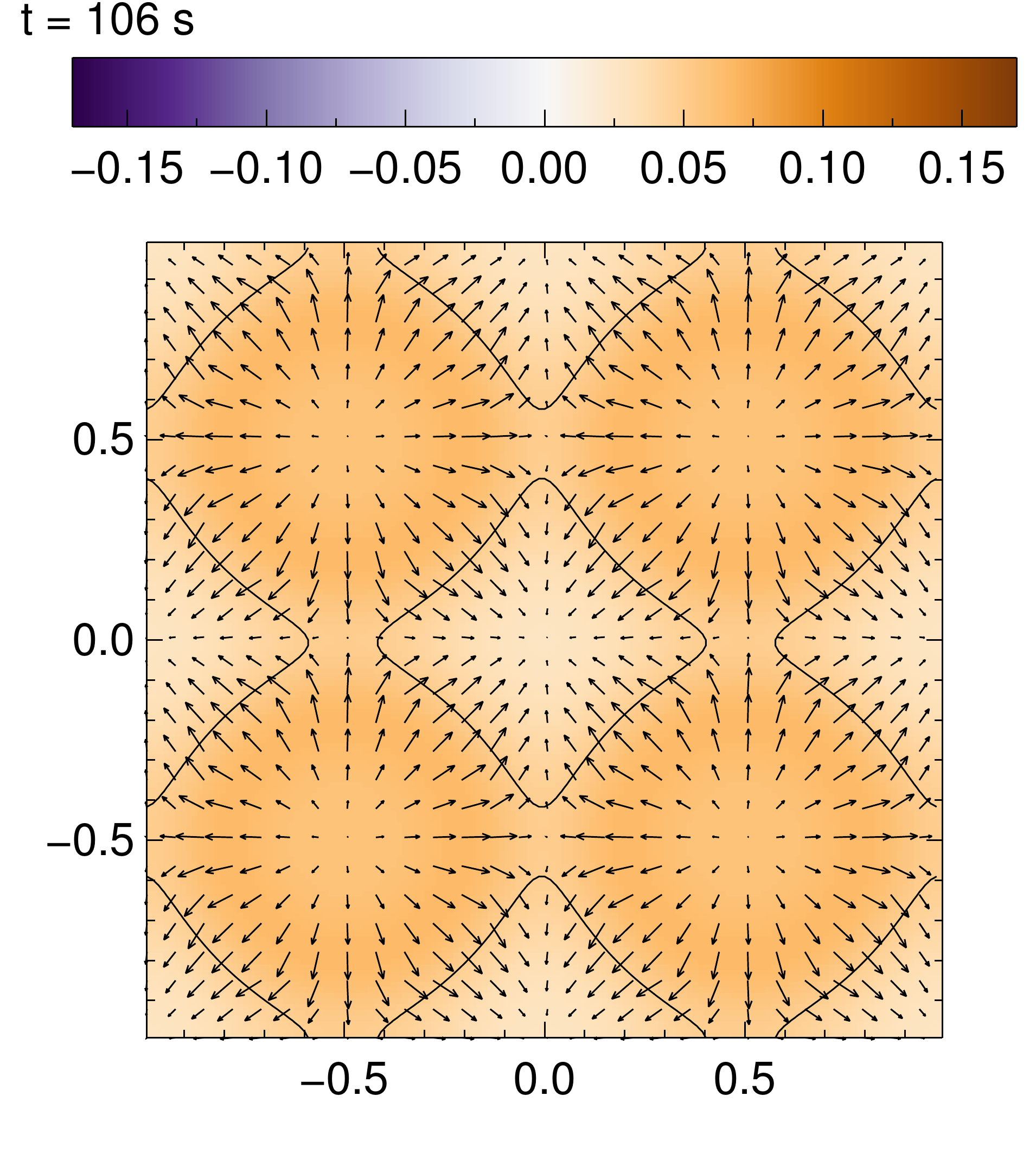}
\caption{Vectors of the horizontal velocity perturbations at 2 Mm at time $t=106$ s. The color scale shows the longitudinal velocity component in m/s and the black contours show the $\beta=1$ border. The spatial scales are given in Mm. \label{fig:vp_with_vectors}}
\end{figure}

To help visualize the direction of the velocity perturbations we overplot the color scale for the parallel component with vectors showing the horizontal velocity perturbation (Fig. \ref{fig:vp_with_vectors}) for the same time and height as in Fig. \ref{fig:pna_horizontal_0deg}. It is now obvious that the plasma expands from the loop centers toward the centers of the loop exteriors (and half of a period later vice versa). Where the loops are closest to each other, counter-streaming flows get deflected sideways toward the centers of the loop exteriors. This also solidifies our interpretation of a deformed sausage wave as mode identification.

\begin{figure}
\centering
\includegraphics[scale=0.3]{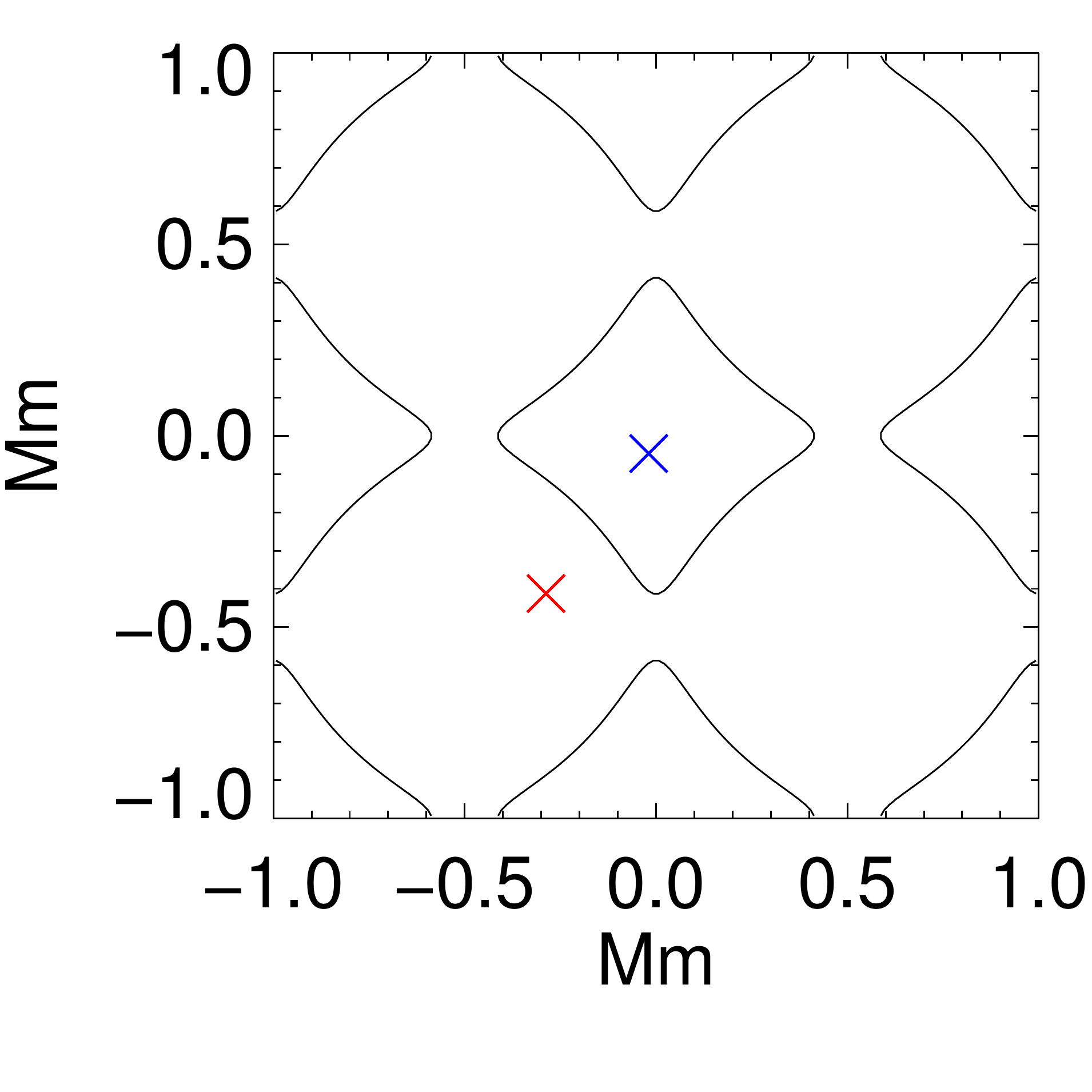}
\caption{Location of the vertical lines of Fig. \ref{fig:wave_propagation_vertical}. The black lines show the $\beta=1$ contour at a height of 2 Mm, where there is high $\beta$ inside and low $\beta$ outside the ``circles''. The red and blue crosses denote the location of the vertical line in the loop interior (Fig. \ref{fig:wave_propagation_vertical} left) and the loop exterior (Fig. \ref{fig:wave_propagation_vertical} right), respectively. \label{fig:line_location}}
\end{figure}

\begin{figure*}
\includegraphics[width=0.5\textwidth]{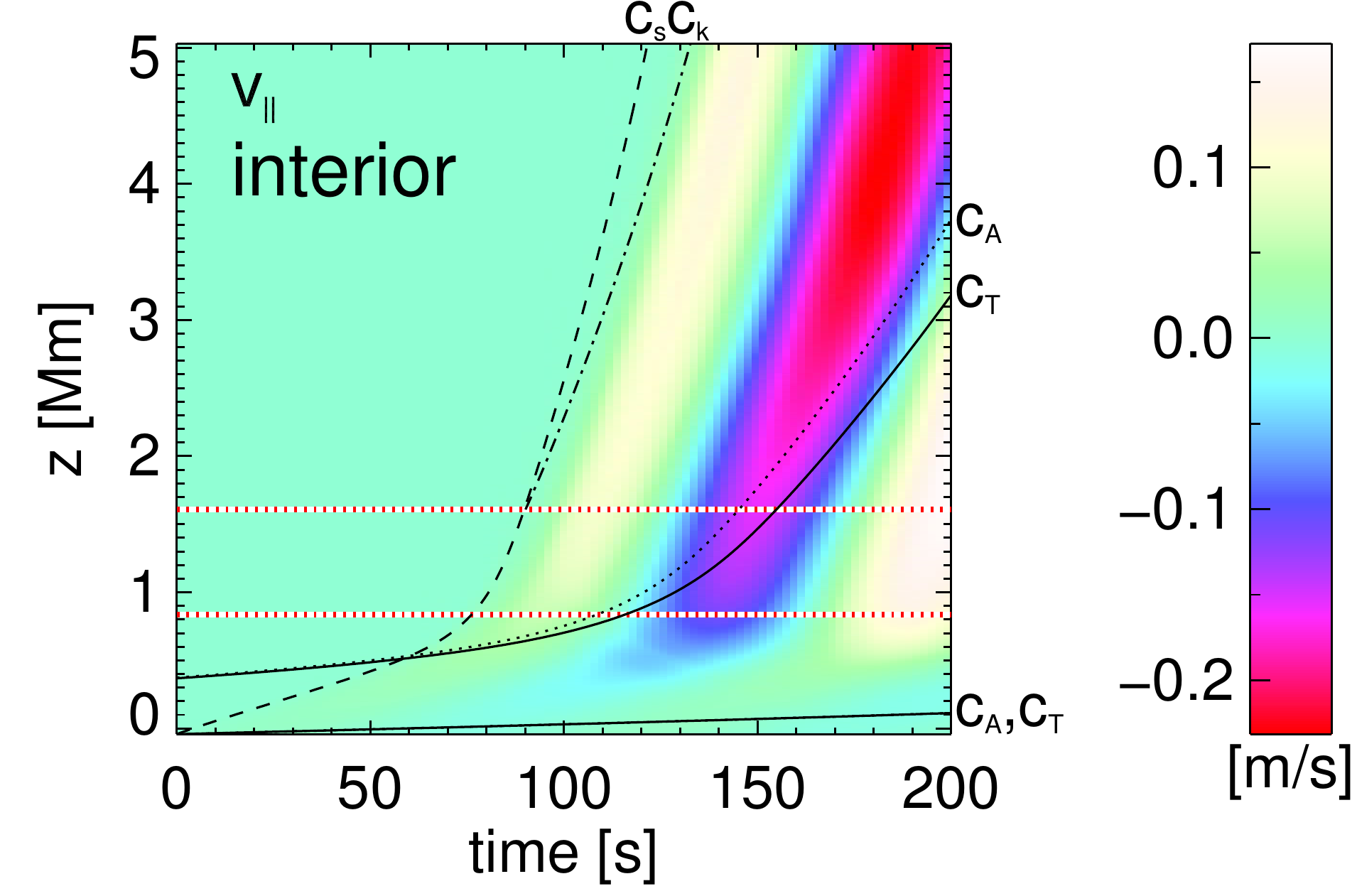} \includegraphics[width=0.5\textwidth]{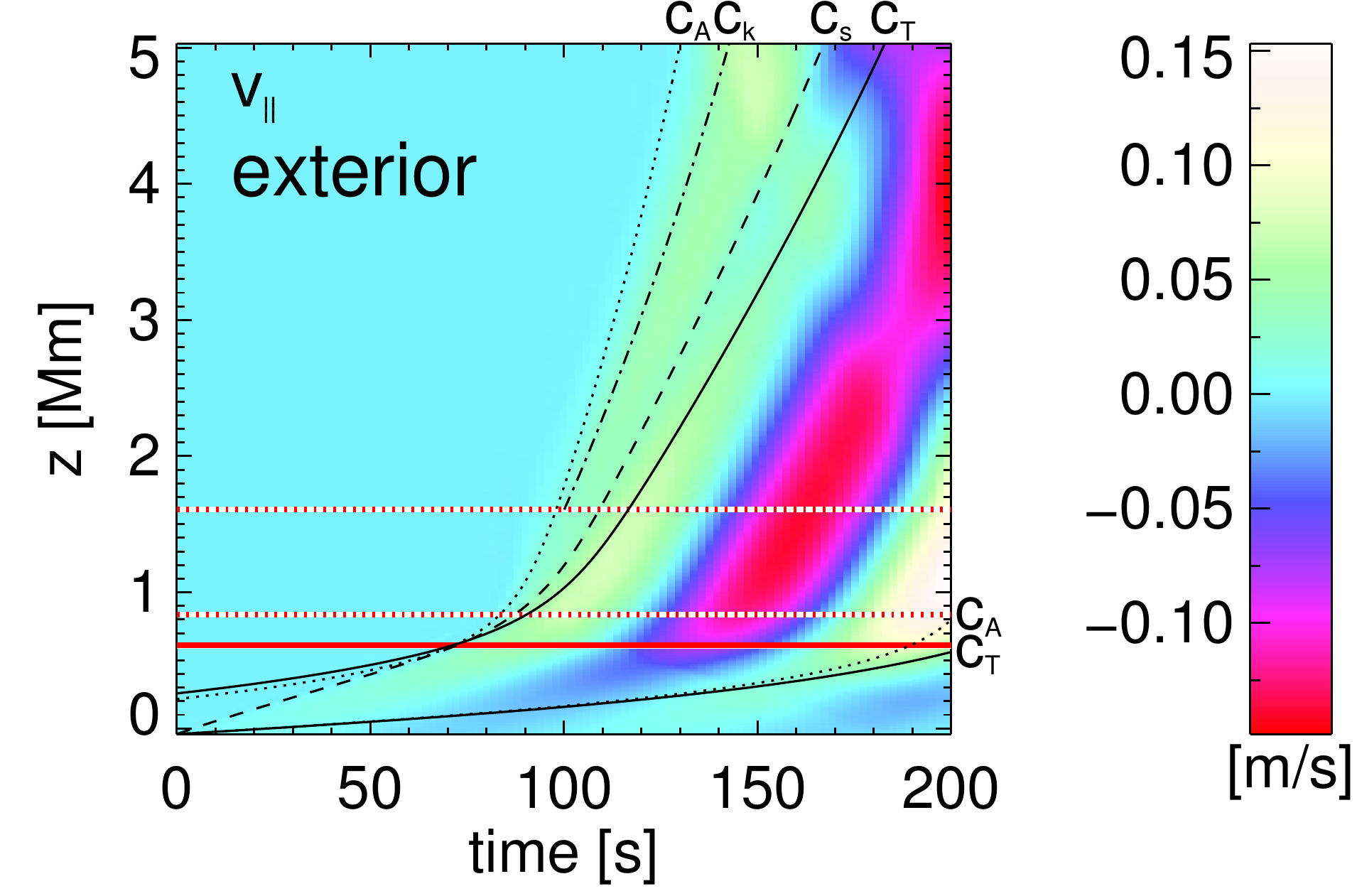}
\includegraphics[width=0.5\textwidth]{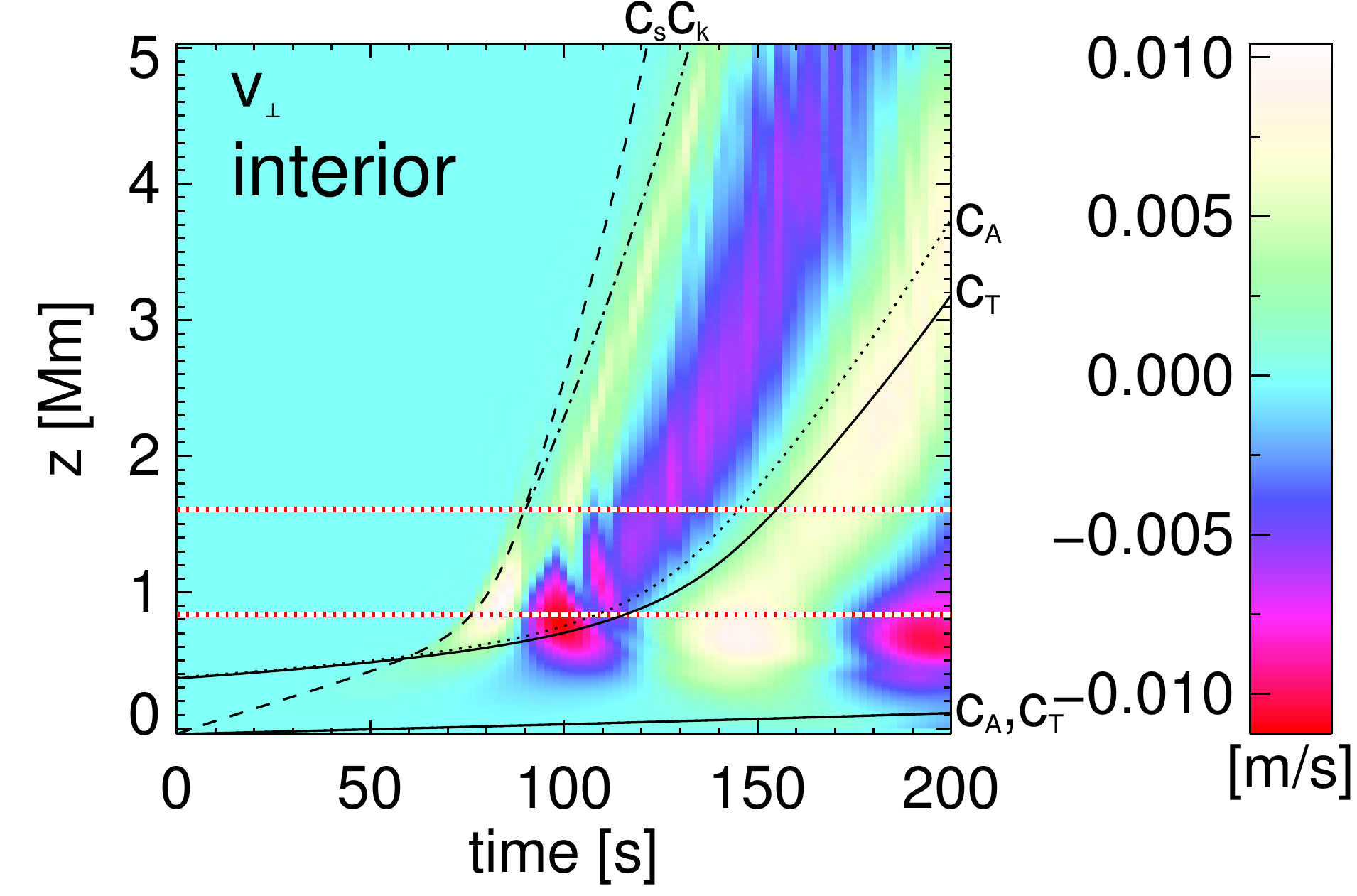} \includegraphics[width=0.5\textwidth]{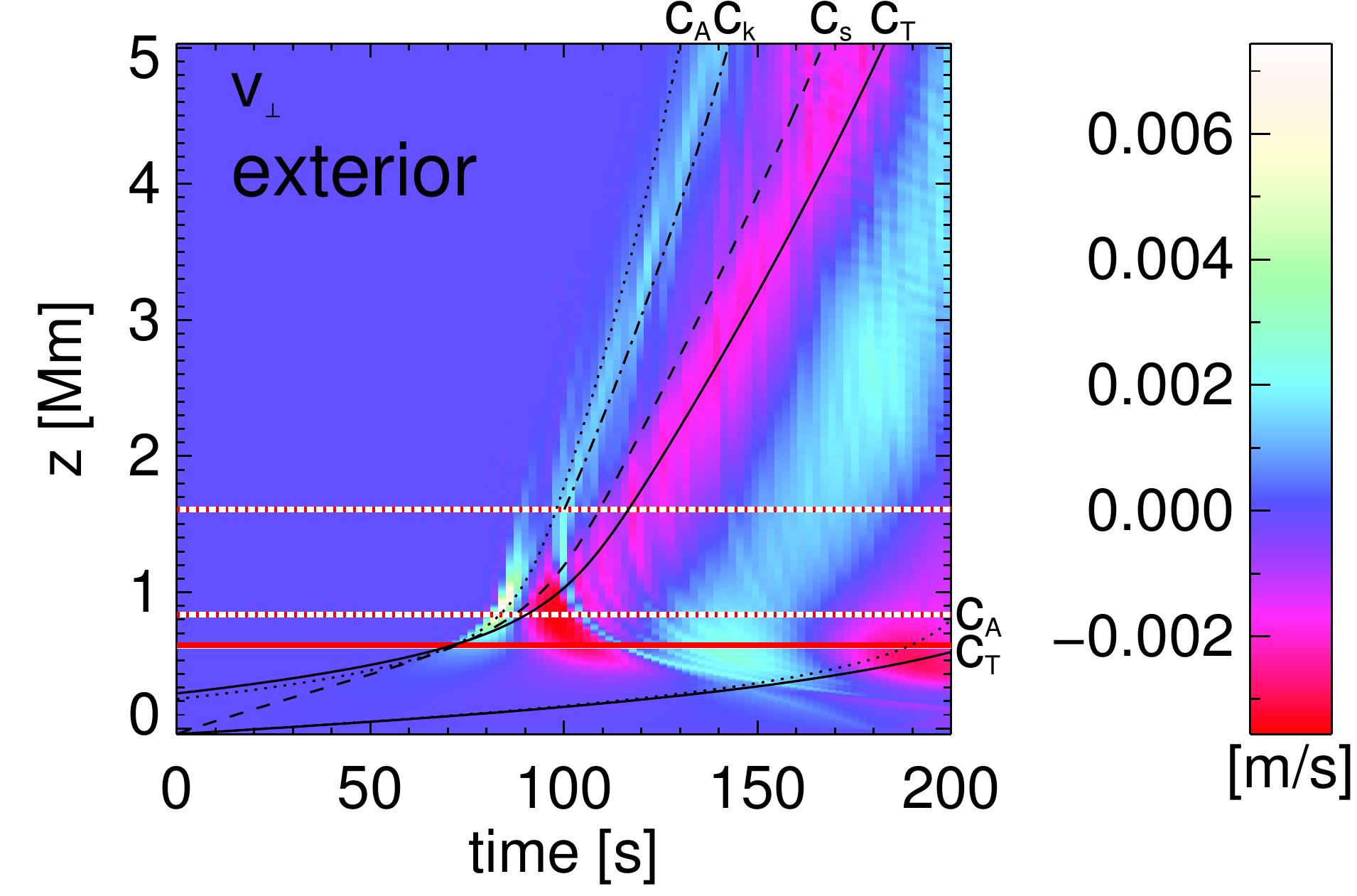}
\includegraphics[width=0.5\textwidth]{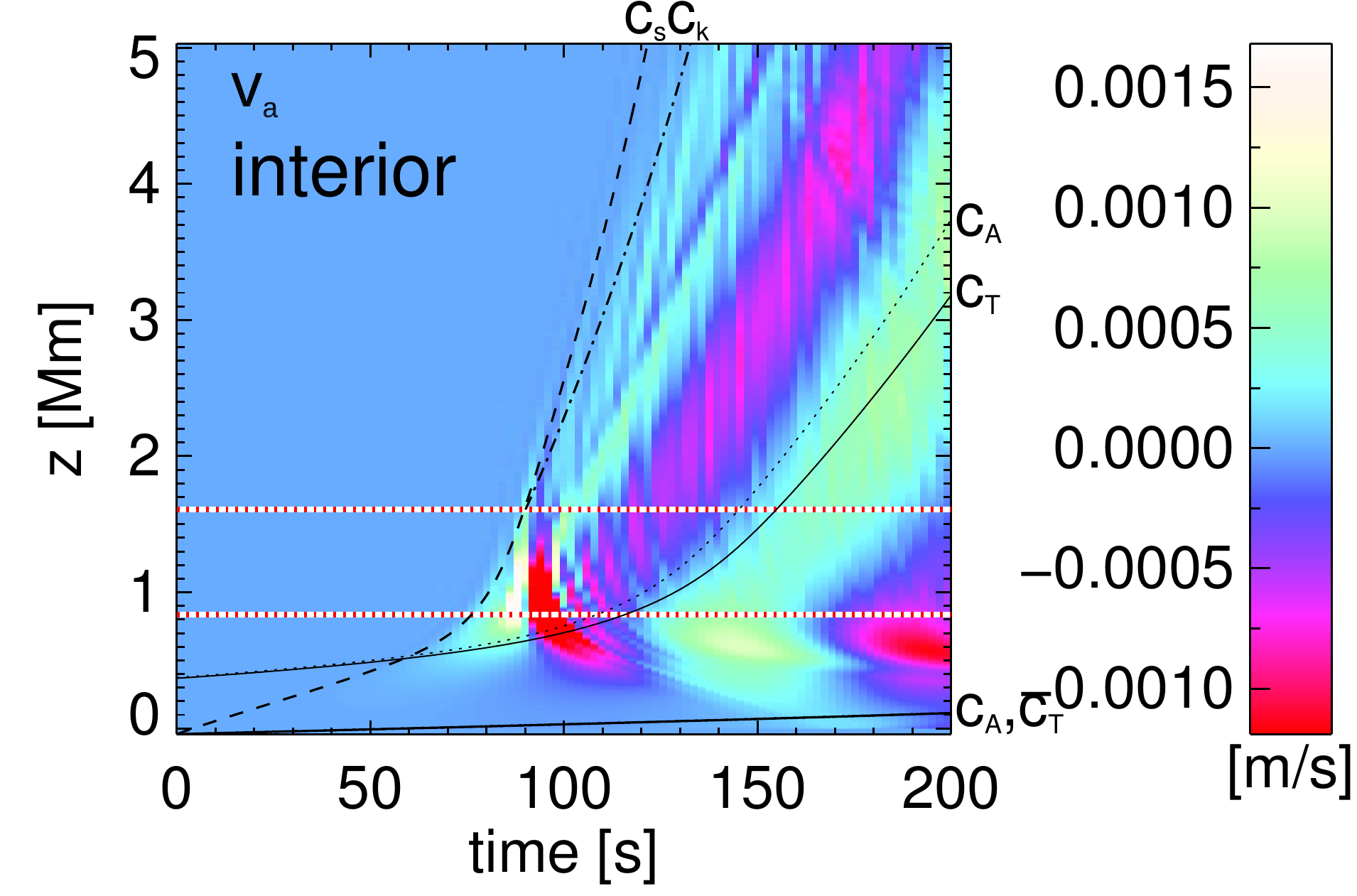} \includegraphics[width=0.5\textwidth]{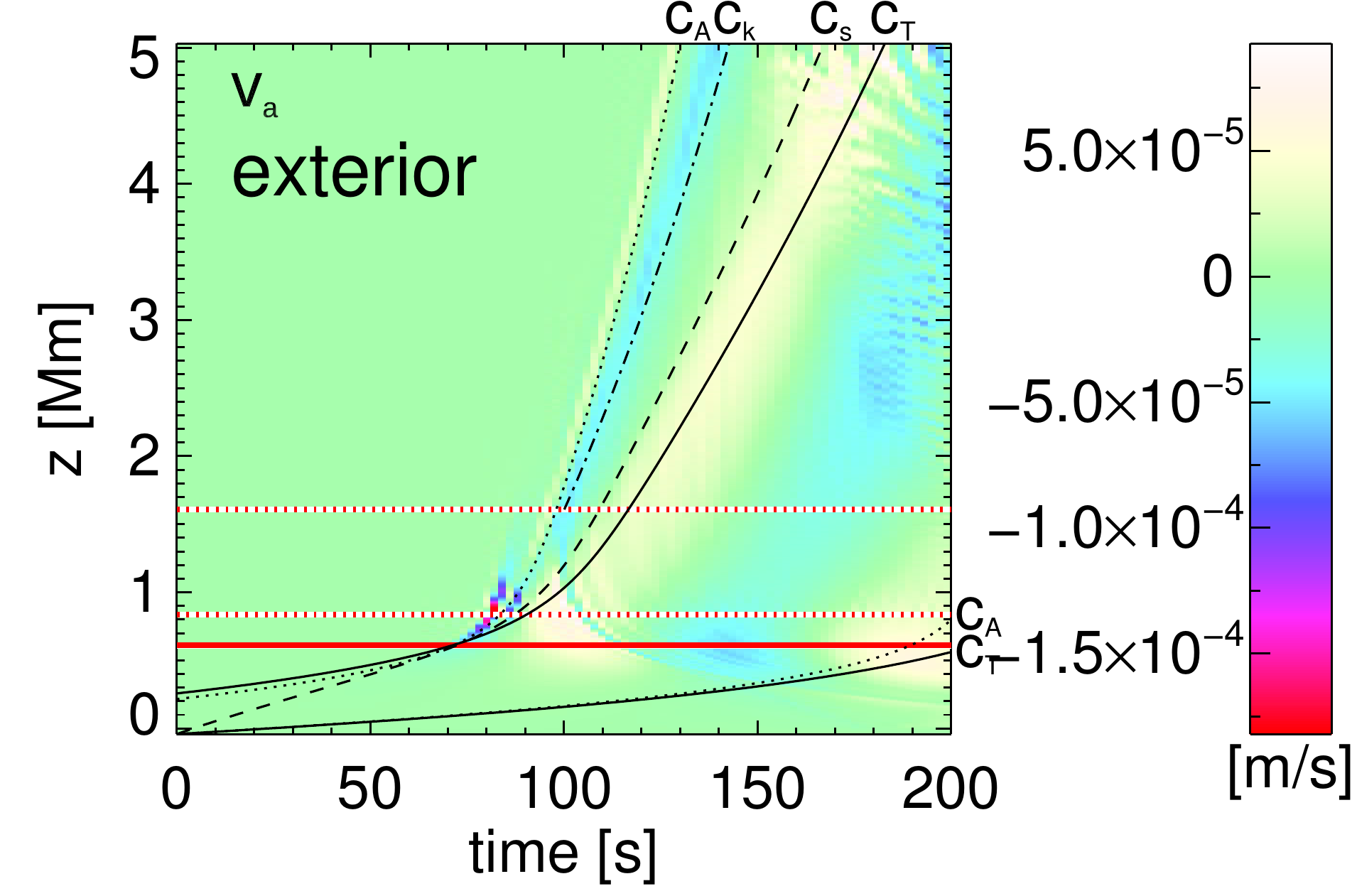}
\caption{Time-distance plot for the longitudinal (top), normal (middle), and azimuthal (bottom) velocity component for a vertical line in the loop interior (left) and loop exterior (right). The dotted red lines denote the transition region and the solid red line shows the $\beta=1$ height with high $\beta$ beneath the line. Plots without that line have high $\beta$ for all heights. The dashed black lines show the sound speed, dotted black lines show the Alfv\'{e}n speed, dash dotted black lines show the kink speed, and solid black lines show the tube speed. The extreme values in the middle right plot are saturated to make the other structures visible as well. \label{fig:wave_propagation_vertical}}
\end{figure*}

In order to study the wave propagation through the system, we looked at the data of two vertical lines in the domain: one of the lines is located in the loop interior where there is a low magnetic field, while the other is close to the center of the loop exterior, which has the maximum magnetic field. The vertical line in the loop interior is at a distance of $0.69R$ from the loop center and therefore at the same position in our MHS atmosphere as the ratio of displacements in Fig. \ref{fig:displacement_ratio} (left) in the simplified model of Sect. \ref{subsec:expected_modes}. We refrain from using data from the exact center of the loops and loop exteriors because the azimuthal unity vector $\vec{\hat{e}}_a$, and therefore also for the normal unity vector $\vec{\hat{e}}_\perp$, are not defined there. The positions of the vertical lines are shown in Fig. \ref{fig:line_location}, while the velocity components in these lines are plotted as a function of time in Fig. \ref{fig:wave_propagation_vertical}. In the latter figure, four characteristic speeds are indicated: the local sound speed (dashed black line), local Alfv\'{e}n speed (dotted black line), local tube speed (solid black line), and kink speed (dashed dotted line). As seen in Equation \ref{eq:kink_speed}, the kink speed is calculated by external and internal Alfv\'{e}n speed and density. Those values are the mean of the values inside and outside the loop for each height with the border at $\beta=1$. Since the loop width is very constant above the transition region, the kink speed is only plotted from the transition region upward. 

The plots on the left side of Fig. \ref{fig:wave_propagation_vertical} show the velocity components in the loop interior, where there is high $\beta$ for all heights. For the longitudinal component inside the loop (top left) the waves propagate smoothly with approximately the sound speed or kink speed and do not seem to be disturbed by the transition region (region between red dotted lines), except for some slight reflection, which is visible between 110 to 130 seconds. No features propagating with slower speeds are visible. However, if we look onto the same component but in the loop exterior (Fig. \ref{fig:wave_propagation_vertical} top right), we see a different picture. There, the waves have to travel through the $\beta=1$ layer before going through the transition region. Below this border, where $\beta \gg 1$, the waves travel again with the sound speed as for the loop interior, but as soon as the first waves pass the $\beta=1$ layer there are suddenly wave features that travel with the Alfv\'{e}n or tube speed. Following the definition for mode conversion and transition of \cite{cally_2005}, this could be interpreted as a conversion from fast acoustic waves to fast magnetic waves. However, we have to be cautious when calling a wave acoustic or magnetic above the transition region in our model because we are dealing with tube waves with high $\beta$ inside the loops and low $\beta$ outside the loops. In addition, we also see features propagating with the sound speed or tube speed above the $\beta=1$ layer, which seems like a transition from fast acoustic waves to slow acoustic waves at first sight. Similar to the waves inside the loop, there is a sign of reflection from the transition region, which is best visible at a time of about 135 seconds. However, we also have unwanted reflections from the upper boundary, which distort the wave shapes coming from below.

Compared to the longitudinal velocity components, the normal velocity components (Fig. \ref{fig:wave_propagation_vertical} middle) are about one order of magnitude smaller. This is no surprise, as only the longitudinal component is driven at the bottom of the domain, while the other components arise from mode conversion or coupling due to inhomogeneity. The general behavior is similar to the longitudinal component, where there are only waves propagating with the sound speed or kink speed in the loop interior and a combination of slow and fast waves in the loop exterior. However, apart from the much stronger reflections from the transition region compared to the general amplitude and the much less prominent reflections from the upper boundary, two striking effects appear. The first is the high velocity amplitude around the lower border of the transition region, which we do not investigate in this paper. From the first high amplitude wave, a wave is launched that travels faster than the fastest characteristic speed (i.e., sound speed for the loop interior and Alfv\'{e}n speed for the loop exterior) until it vanishes. This could be a sign of a leaky sausage wave, as these waves can travel faster than the external Alfv\'{e}n speed \citep{pascoe_etal_2007}, which would be the Alfv\'{e}n speed in the center of the loop exterior for our case. By looking closer at the simulation data we find indications that this is indeed the case. We did not find a leaky sausage mode with high phase speed for our simplified model in Sect. \ref{sec:methods}, however, it could still appear in our more complicated MHS model. The second effect are the vertical stripe patterns, which are more pronounced inside the loop than outside. These may either be interference patterns by partial reflection of the waves due to the temperature gradient (the vertical temperature gradient is stronger inside the loops than outside), or just artifacts due to the cylindrical structure within a Cartesian grid. Those patterns are not of interest for our study and are therefore not considered in the following.

The azimuthal velocity component (Fig. \ref{fig:wave_propagation_vertical} bottom) closely resembles the normal component, except for the generally smaller amplitude and that we now also see waves with a phase speed of approximately the Alfv\'{e}n speed inside the loop. In addition, this component is more strongly affected by the reflection from the upper boundary. The close relation between the normal and azimuthal component is no surprise, as Fig. \ref{fig:vp_with_vectors} shows us that the azimuthal velocities arise from the deflection of expanding (normal) plasma movements toward the centers of the loop exteriors.

To roughly estimate how much of the wave energy is reflected at the transition region in Figure \ref{fig:wave_propagation_vertical}, we determine the wave energy flux parallel to the magnetic field and look at the ratio of maximum (positive, upward) to minimum (negative, downward) amplitude of the first wave front. This shows that about 3\% of the flux is reflected in the loop interior and about 4\% in the loop exterior. Because of this crude approximation there might be an error of several percent, but the reflected flux is still very low. For a steeper transition region, more reflection would be expected. We refrain from giving an estimate about the amount of flux reflected from the upper boundary, since this is a purely numerical issue that holds no physical value.

From Fig. \ref{fig:pna_horizontal_0deg} we get an approximate loop radius of $R \approx 0.5$ Mm and from \ref{fig:wave_propagation_vertical} we can estimate that the wavelength lies between 5 and 10 Mm. From those values we determine that $kR$ lies between 0.31 and 0.63, which allows us to compare our simulation data to the wave mode solutions of the simplified model in Figs. \ref{fig:phase_speed_diagram} and \ref{fig:displacement_ratio} for small $kR$. All our data suggests that the parallel displacement is larger than the perpendicular displacement, which immediately excludes the fast kink surface mode. We also exclude the slow kink body mode because there is no reason why a kink mode would be excited owing to the symmetry of our system. We therefore conclude that the fast waves we see in Fig. \ref{fig:wave_propagation_vertical} are fast sausage surface waves, which is supported by the fact that the magnitude of the velocity perturbations is smaller in the loop center than in its surroundings, while the slow waves are slow sausage body waves. However, for small $kR$ there is no solution of fast sausage surface waves (Fig. \ref{fig:phase_speed_diagram}), so the simple model we used to calculate the phase speed diagram does not describe those wave modes well in that region. In addition, we might still have a plane-like wave traveling through the whole domain, which mostly ignores the horizontal structuring because of its relatively slow changes. It is, however, difficult to distinguish such waves from the fast surface sausage waves.

\subsection{Wave propagation for inclined flux tubes}

\begin{figure*}
\includegraphics[width=\textwidth]{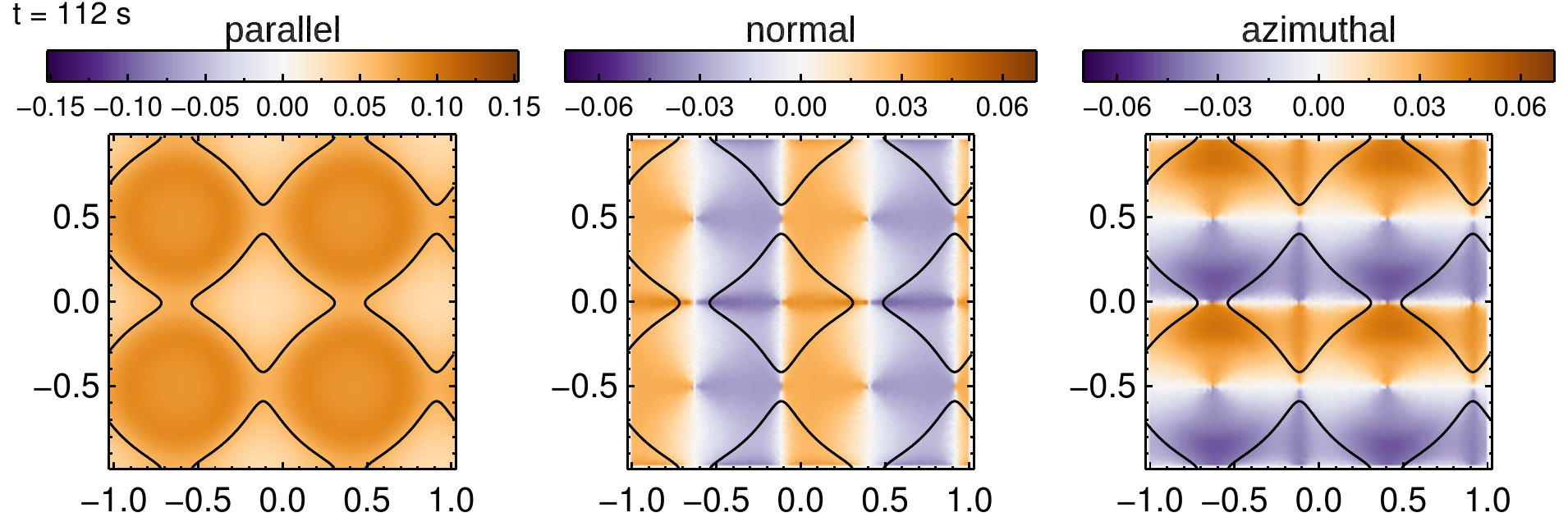}
\caption{Components of the velocity perturbation in a horizontal cut at 2 Mm for $\theta=15^{\circ}$ at a time of 112 s after the start of the simulation. The velocities are given in m/s and the spatial scales are in Mm. The black pseudo-circles show the $\beta=1$ contour, with $\beta\gg1$ inside and $\beta<1$ outside the loop. The values of the first two pixels next to the margin are set to zero for the normal and azimuthal component, as the corresponding vectors were badly defined in that region. The temporal evolution is available as an online movie. \label{fig:pna_horizontal_15deg}}
\end{figure*}

We now study the case with the inclined magnetic field for an inclination of $\theta = 15^{\circ}$. The time development of a horizontal cut of the simulation box with $15^{\circ}$ inclined magnetic field at a height of 2 Mm is shown in the movie that is available online. A snapshot of this movie at $t=112$ s is shown in Fig. \ref{fig:pna_horizontal_15deg}. 
The parallel velocity component in Fig. \ref{fig:pna_horizontal_15deg} behaves very similarly to the vertical case (Fig. \ref{fig:pna_horizontal_0deg}), where there are higher velocity perturbations within the loops than outside, which change signs over time. However, both the normal and azimuthal components show not only a similar stripe pattern, but also have the same magnitude. These effects arise from the whole plasma moving first to the left (like in the figure) and later to the right\footnote{The vector normal to the flux surface $\vec{\hat{e}_\perp}$ does also have a component in the $z$-direction for $\theta \neq 0$, so strictly speaking the plasma also moves up and down.}, which corresponds to a kink wave. During the transition from plasma moving from one direction to the other, we have a short time span in which plasma flows both to the left and right. Another
kind of fast wave was also excited because we expect the magnitude of perpendicular displacement to be larger than the magnitude of parallel displacement for fast kink surface waves (see Fig. \ref{fig:displacement_ratio}); however, we find the opposite in Fig. \ref{fig:pna_horizontal_15deg}. This wave is either a fast sausage surface wave, such as in the case with vertical flux tubes or a plane wave that does not ``feel'' the loop structure.

\begin{figure}[]
\centering
\includegraphics[scale=0.3]{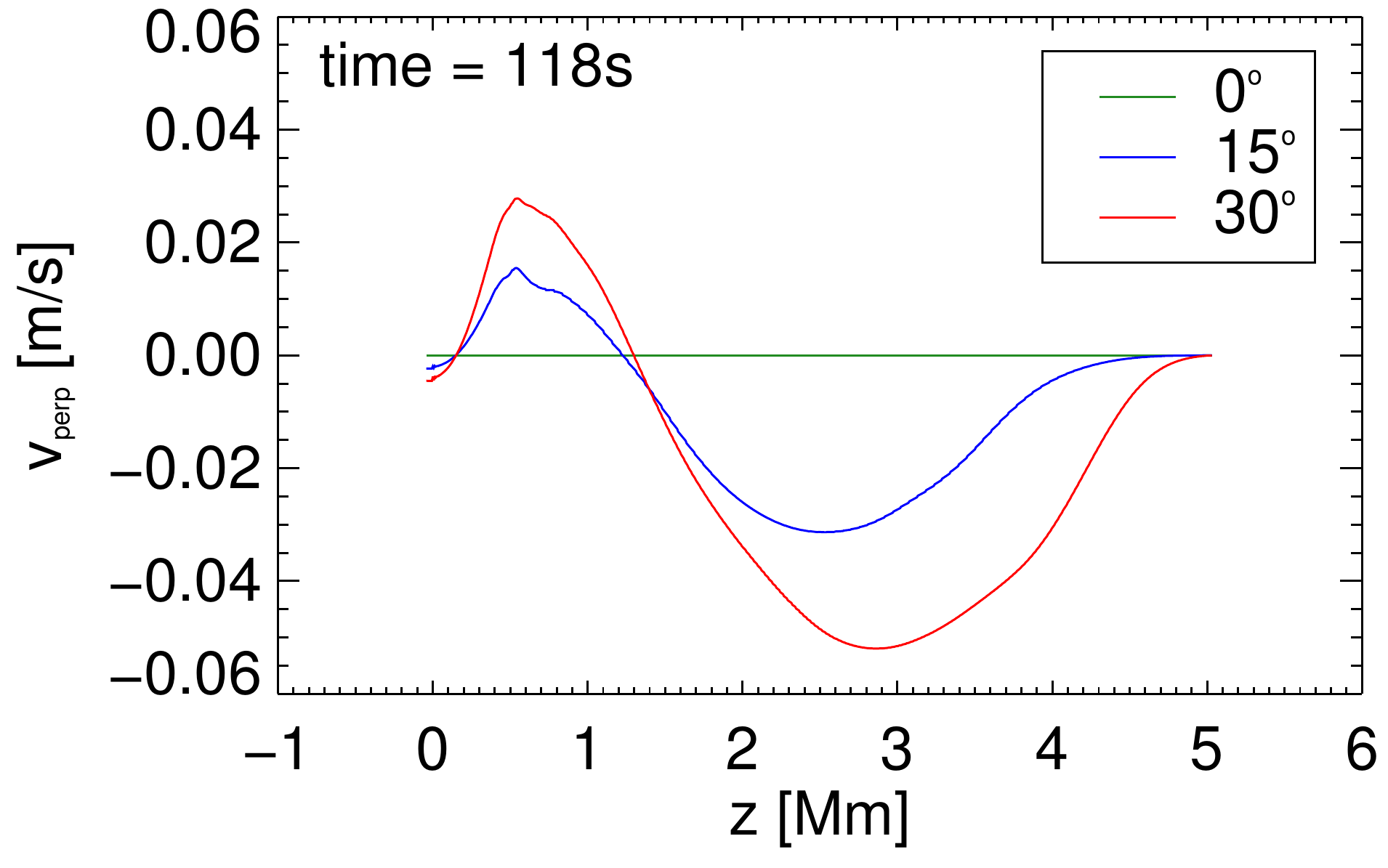}
\caption{Velocity perturbation perpendicular to the magnetic field in the loop center as a function of height at time $t=118$ s for three different magnetic field inclinations.\label{fig:kink_oscillation}} 
\end{figure}

We checked the kink wave assumption by plotting the velocity disturbance perpendicular to the magnetic field in the loop center as a function of the height, where we find upward propagating waves for the inclined cases with $\theta=15^{\circ}$ and $\theta=30^{\circ}$, whereas there is no kink wave for the vertical case ($\theta=0^{\circ}$). These kink waves are excited in the simulations with the inclined loops, as the driver is still purely vertical and therefore gives the loops a push perpendicular to the magnetic field. Since the magnetic field is only inclined in the $x$-direction, there is no significant perturbation of the loop centers in the $y$-direction. A snapshot of these waves is shown in Fig. \ref{fig:kink_oscillation}. 

\begin{figure*}
\includegraphics[width=0.5\textwidth]{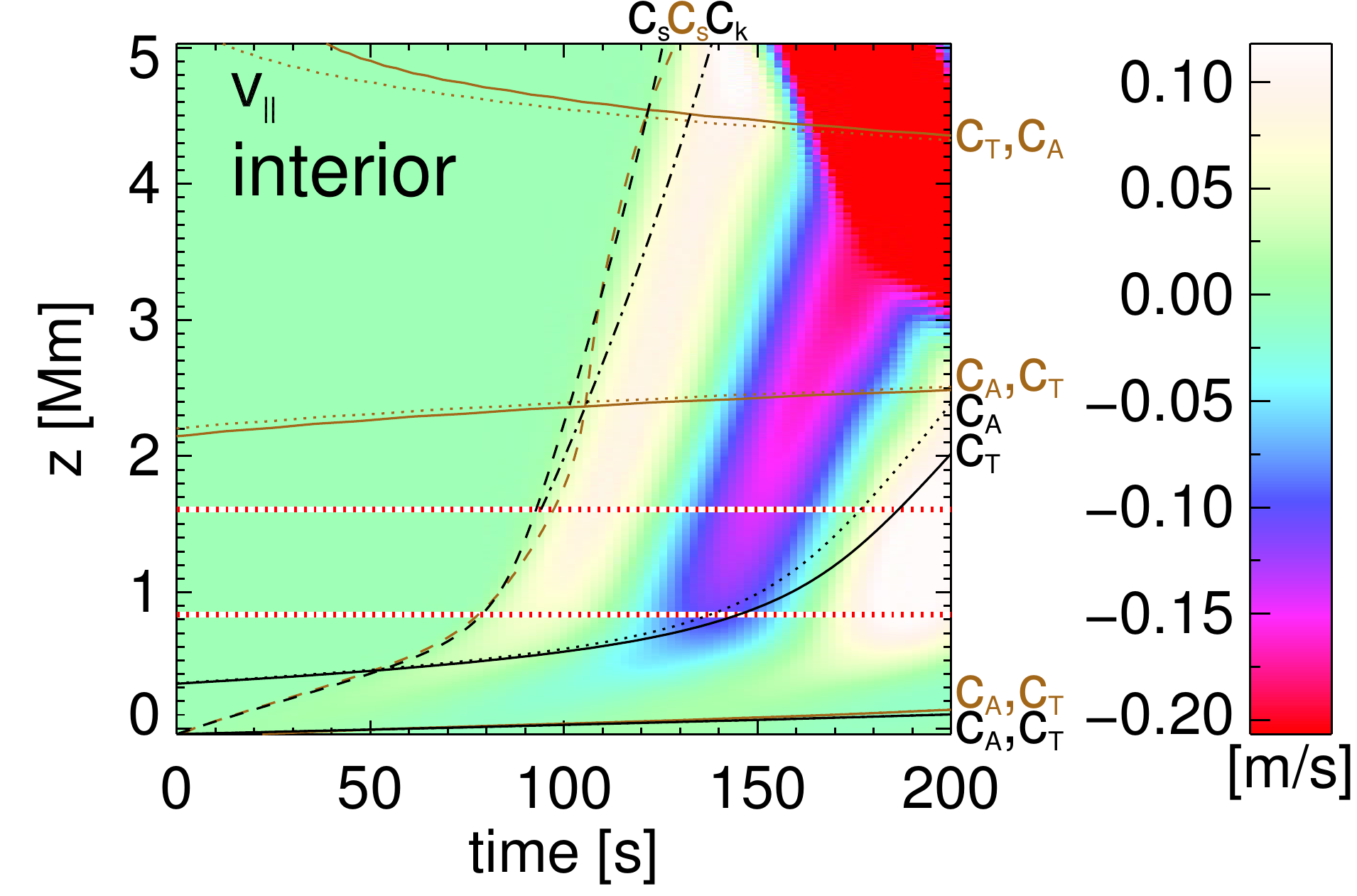}
\includegraphics[width=0.5\textwidth]{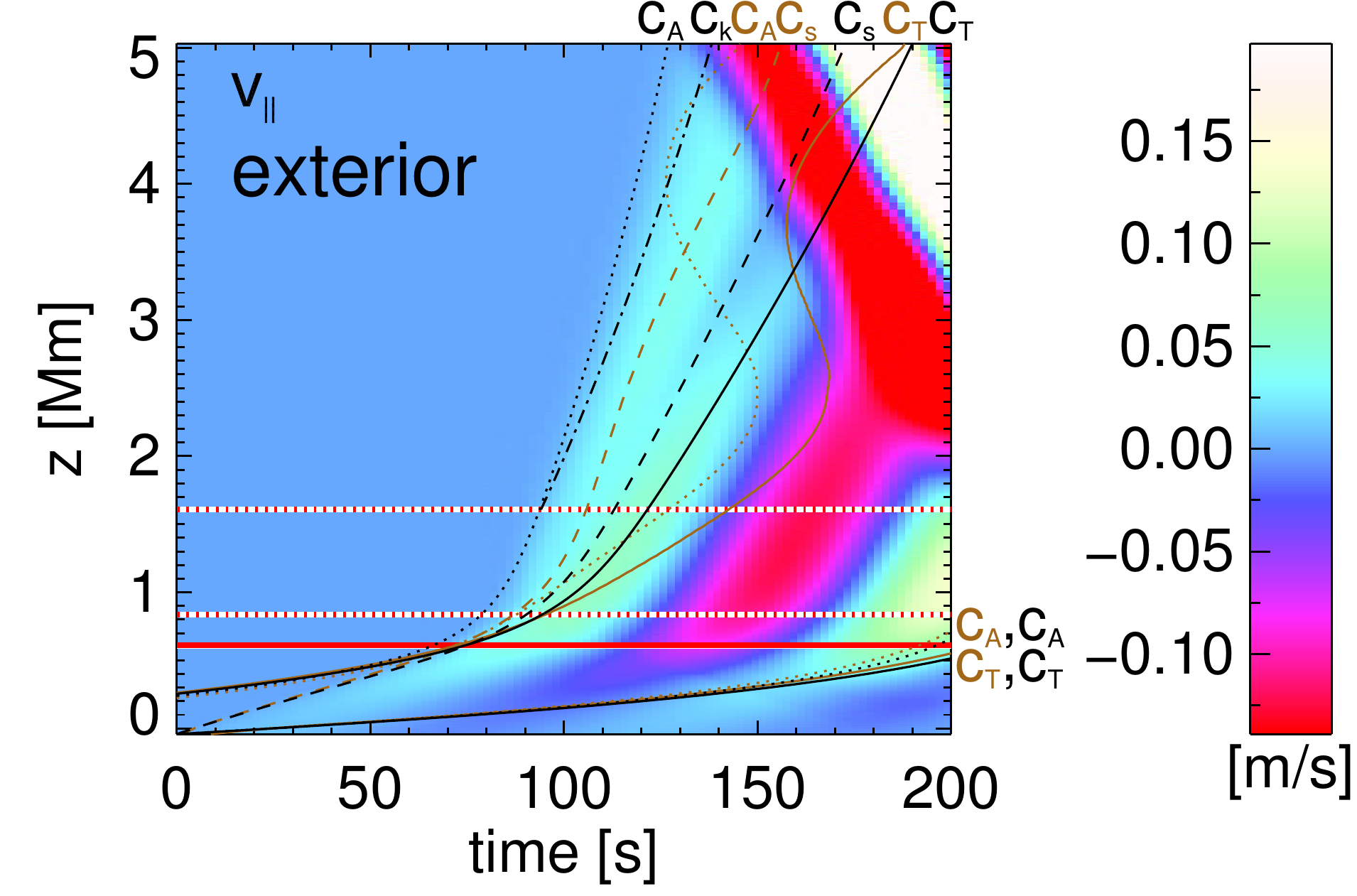}
\includegraphics[width=0.5\textwidth]{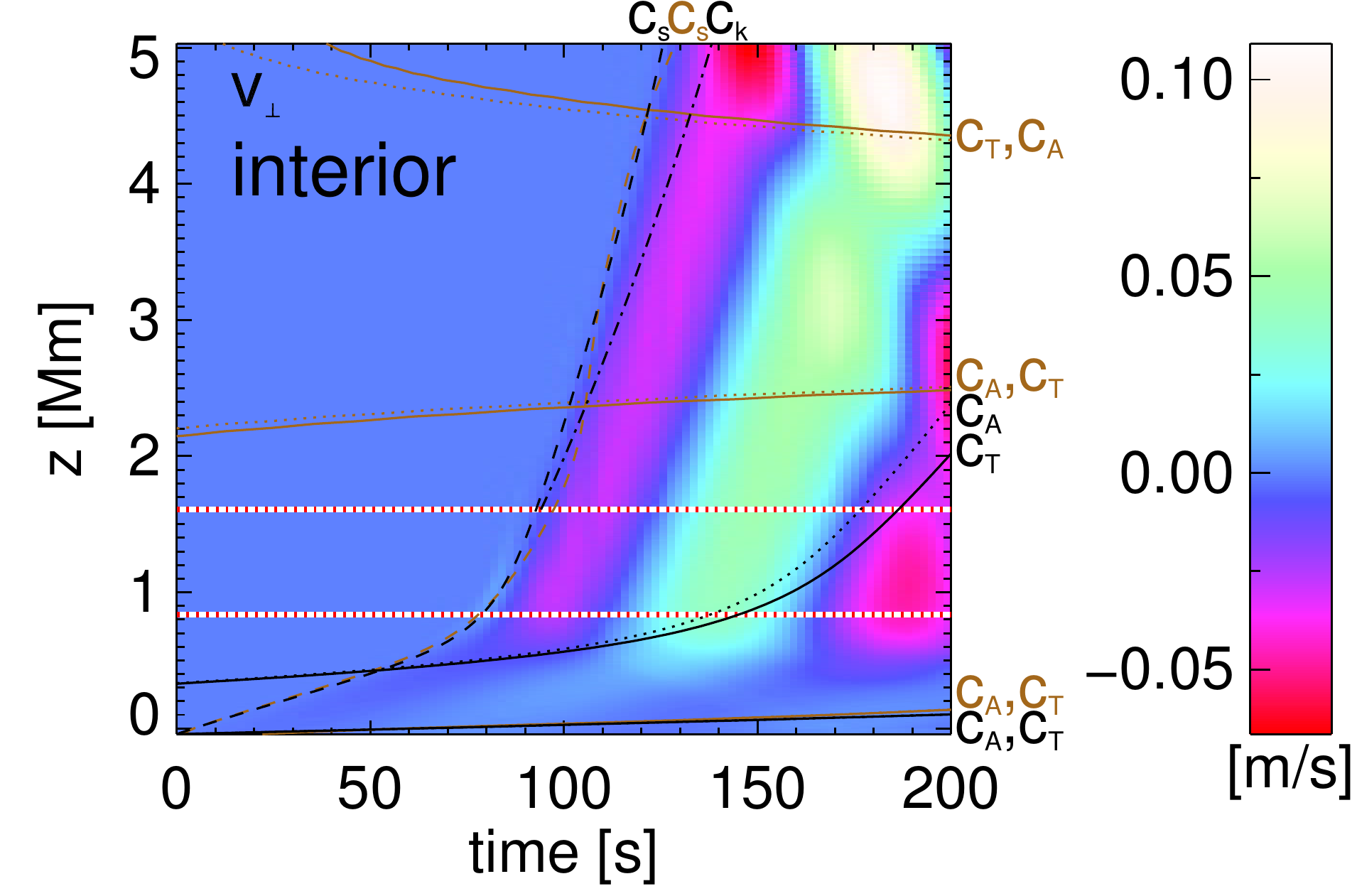}
\includegraphics[width=0.5\textwidth]{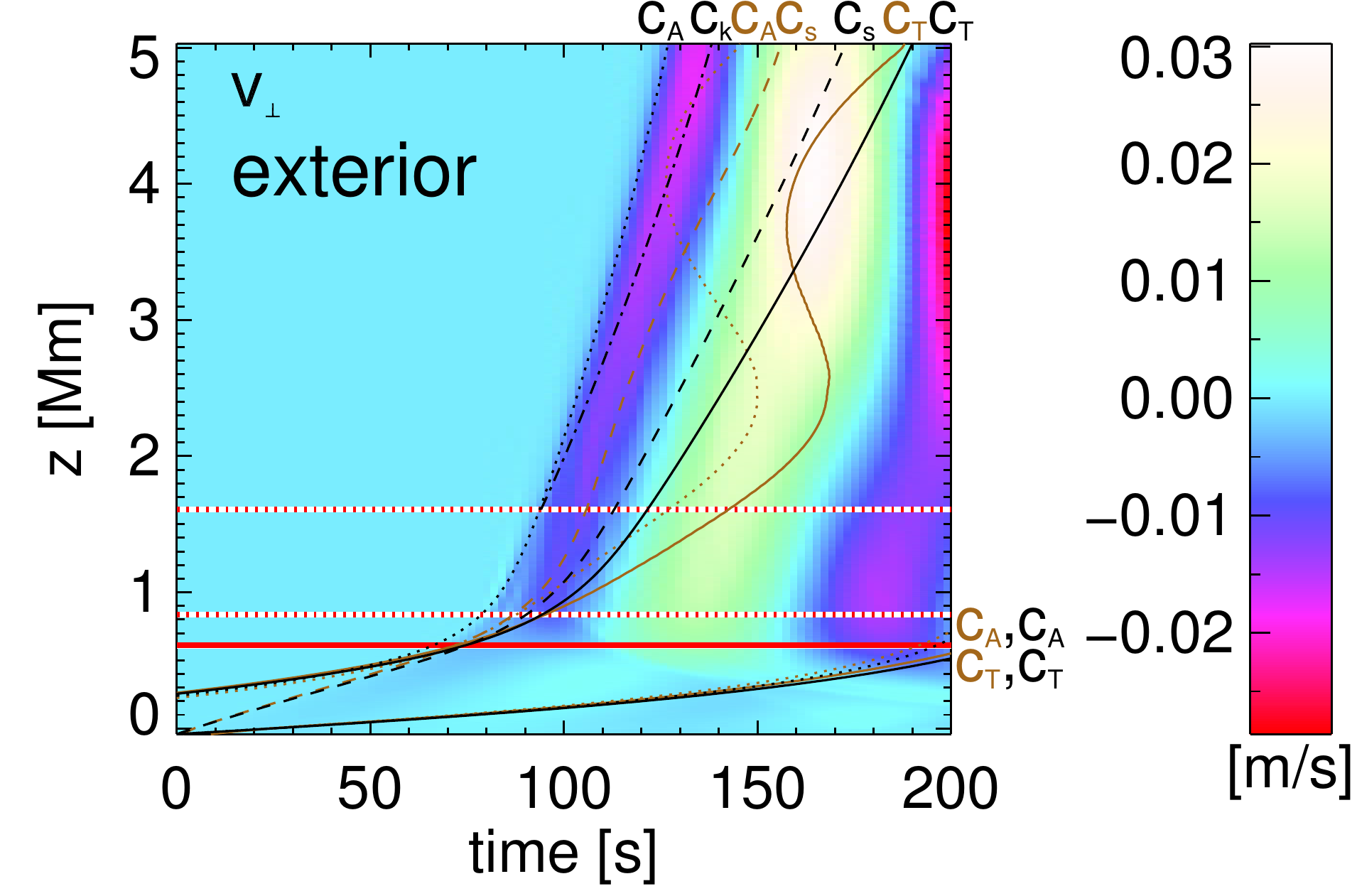}
\includegraphics[width=0.5\textwidth]{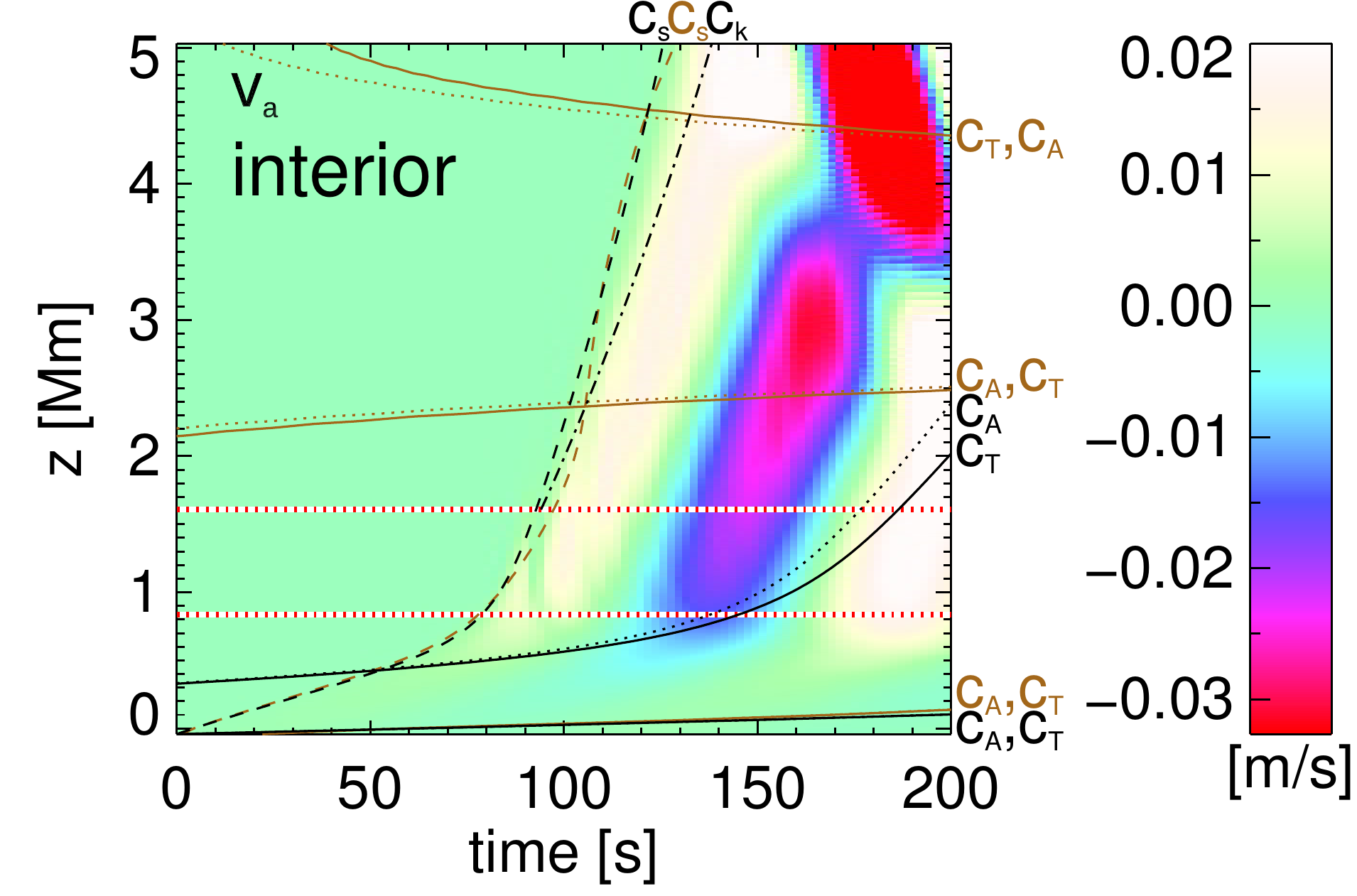}
\includegraphics[width=0.5\textwidth]{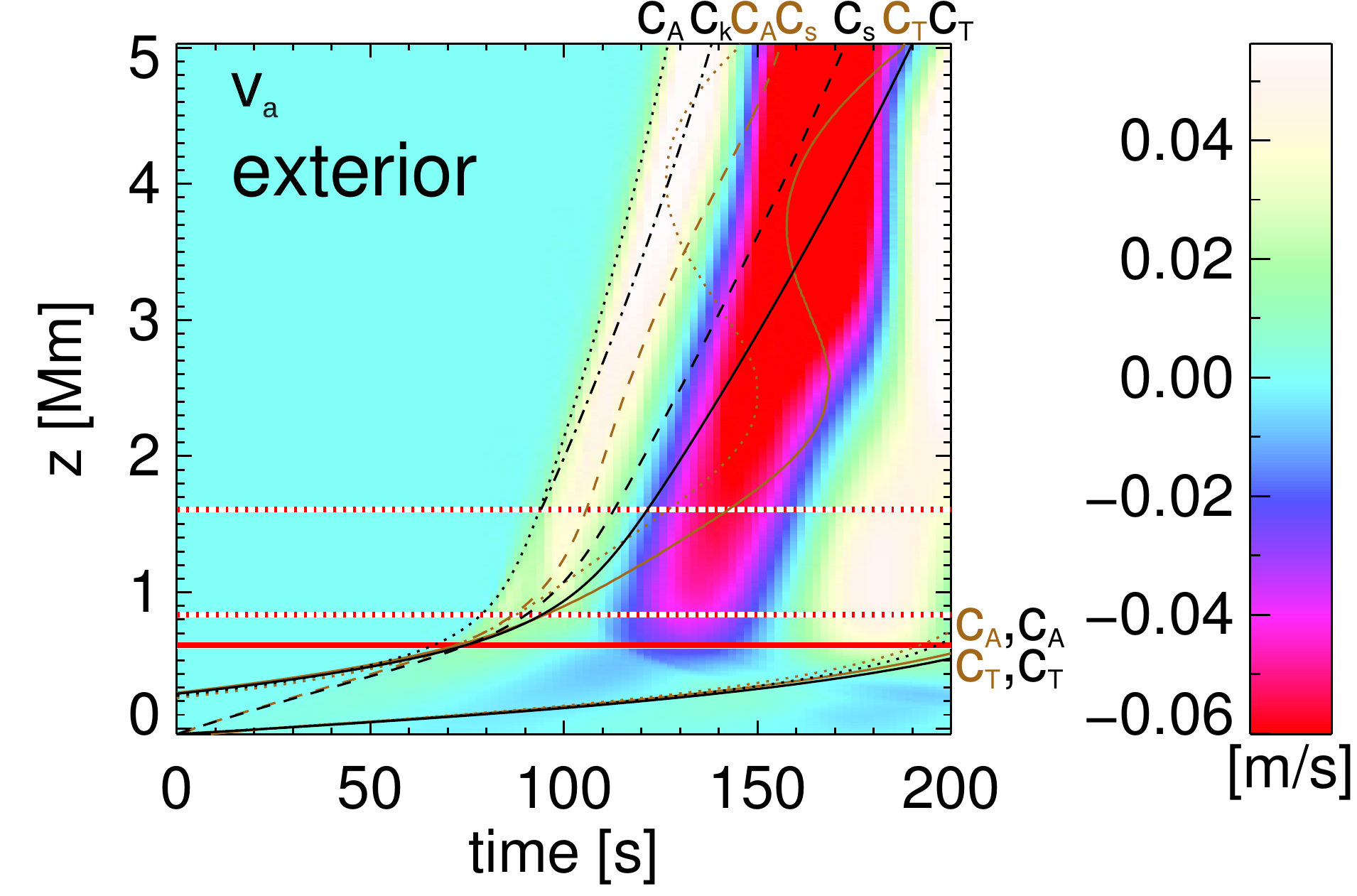}
\caption{Time-distance plot for the parallel (top), normal (middle), and azimuthal (bottom) velocity component for a line in the loop interior (left) and loop exterior (right) in a simulation with a magnetic field that is inclined $15^\circ$. The dotted red lines indicate the transition region and the solid red line shows the $\beta=1$ height with high $\beta$ beneath the line. The plots without that line have high $\beta$ for all heights. The dashed lines show the sound speed, dotted lines show the Alfv\'{e}n speed, and solid lines show the cusp speed. The black lines show these speeds for wave propagation along the magnetic field, while the brown lines show speeds for vertical wave propagation. The extreme values (only in reflection zones) in the plots for the parallel and azimuthal component are saturated to make the other structures visible as well. \label{fig:wave_propagation_inclined}}
\end{figure*}

Figure \ref{fig:wave_propagation_inclined} shows the equivalent of Fig. \ref{fig:wave_propagation_vertical} for the inclined case, for which we examine the wave propagation along two lines inclined $15^{\circ}$ from the vertical (parallel to the magnetic field) that lie within the loop interior and loop exterior, respectively, at equivalent locations as for the vertical case. The black lines in Fig. \ref{fig:wave_propagation_inclined} show the local sound speed (dashed), local Alfv\'{e}n speed (dotted), local tube speed (solid), and kink speed (dashed dotted) for wave propagation along the loop, i.e., along the magnetic field. Similarly, the brown lines show these speeds for vertical wave propagation, i.e., the direction of the driver. The reason for the strange S-shape for the vertical propagation speeds is that for different heights $z$ vertically propagating waves that reach that height also start from different horizontal locations $x$, where the characteristic speeds are different. Therefore, for some vertical lines, the waves would find more ``favorable'' conditions to propagate than for others, which causes them to arrive earlier at a larger height. The S-shape is more pronounced for the Alfv\'{e}n speed (and therefore also the tube speed) because the Alfv\'{e}n speed changes more between loop interior and exterior than the sound speed. Since vertically propagating waves would have to go through regions with very low Alfv\'{e}n speed for the $y$ location of the loop-interior-line (left column), the S-shape is extreme in those plots. However, we do not see any signs in Fig. \ref{fig:wave_propagation_inclined} of waves propagating along these strange (brown) lines. Therefore, we can conclude that the waves mainly propagate along the magnetic field (black lines) from the transition region onward.

From Fig. \ref{fig:wave_propagation_inclined} it is immediately apparent that the reflection problem from the upper boundary is much stronger for the inclined case. For this reason the plots for the parallel component (top row) and for the azimuthal component (bottom row) are saturated to allow a better visibility of the waves propagating upward. What is also noticeable is that the magnitudes of the normal and azimuthal components are much higher than before, as also seen in Fig. \ref{fig:pna_horizontal_15deg}.

As in the vertical case, the parallel component propagates with the sound speed or kink speed in the loop interior for all heights and splits at the $\beta=1$ layer into waves propagating with the sound speed or kink speed and waves propagating with the Alfv\'{e}n speed or tube speed in the loop exterior. In the plots for the normal component we again see (small) wave signatures that propagate faster than the local fast speed in the transition region at about $t=180$ s, which could be a sign of leaky sausage waves. These wave signatures are also present in the azimuthal component, but not visible in the displayed data range. In both the normal and azimuthal component we only see wave propagation with the sound speed or kink speed for high $\beta$ and the Alfv\'{e}n speed or kink speed for low $\beta$.

There is again some wave reflection from the transition region, which is about 3\% of the energy flux in the loop interior and 6\% in the loop exterior. Given that this is just an estimation, it is not possible to say if the reflection is dependent on the inclination angle, since these values are very similar to those obtained for the vertical case (3\% and 4\%).

Similar as before in the vertical case, we compare our simulation results with Figs. \ref{fig:phase_speed_diagram} and \ref{fig:displacement_ratio}. We found from Figs. \ref{fig:pna_horizontal_15deg} and \ref{fig:kink_oscillation} that definitely a kink wave is excited, and from Fig. \ref{fig:wave_propagation_inclined} that it propagates fast. Therefore, we identify it as a fast kink surface wave. To explain the ratios of velocity perturbation components in Fig. \ref{fig:pna_horizontal_15deg} there must also be another fast wave, which could be a fast sausage surface wave or a plane fast wave that ignores the cylindrical structuring. The slow waves appearing in Fig. \ref{fig:wave_propagation_inclined} are symmetric around the center of the loop exterior and are therefore identified as slow sausage body waves.

To reiterate the results of this section, we inserted a fast (acoustic) wave at the bottom of our domain, which converted to different wave modes. This was also found by \cite{cally_2017}, who injected a fast (magnetic) wave into a model with gravitationally stratified Alfv\'{e}n speed profile with discrete (and also ``touching''), inclined flux tubes, using the cold plasma ($\beta=0$) approximation. The initially $m=n=0$ fast waves scattered in Fourier space into other modes, i.e., essentially the $m=0,n=-1$ kink mode. In addition, there was also significant conversion to Alfv\'{e}n waves, which decayed with height as they were also scattered into higher mode numbers as a consequence of mode mixing.

\subsection{Mode conversion} \label{subsec:mode_conversion}

We would like to estimate how much the initially acoustically dominated waves, as excited by the driver, take on magnetic properties. This can be described by mode conversion from acoustic to magnetic waves. When speaking about mode conversion in the following, we mean conversion from acoustic to magnetic behavior, not conversion from fast to slow waves or vice versa. An often used conversion coefficient was given by Equation 26 of \cite{schunker_cally_2006} and we repeat  for convenience, i.e.,
\begin{equation} \label{eq:schunker_conversion}
  C=1-T=1-\exp \Big [ - \frac{\pi k^2 k_\perp^2}{|k_z| (k^2+k_\perp^2)} \Big (\frac{d(c_A^2/c_s^2)}{dz}\Big ) ^{-1} \Big ] _{c_A=c_s} .
\end{equation}
The accuracy of this expression was analytically tested by \cite{hansen_cally_2009}. It is defined by the portion of the wave energy flux that is converted from acoustic to magnetic, where $C=1$ (transmission coefficient $T=0$) describes full conversion from acoustic to magnetic waves (fast wave to fast wave) and $C=0$ ($T=1$) describes no conversion. In this equation, $k$ is the wave number with its components in the $z$-direction $k_z$ and the direction perpendicular to the magnetic field $k_\perp$. The equation is evaluated at the $c_A=c_s$ layer, where mode conversion is supposed to happen. In the following, we check Equation \ref{eq:schunker_conversion} in our simulation data.

\begin{figure}[]
\centering
\includegraphics[scale=0.4]{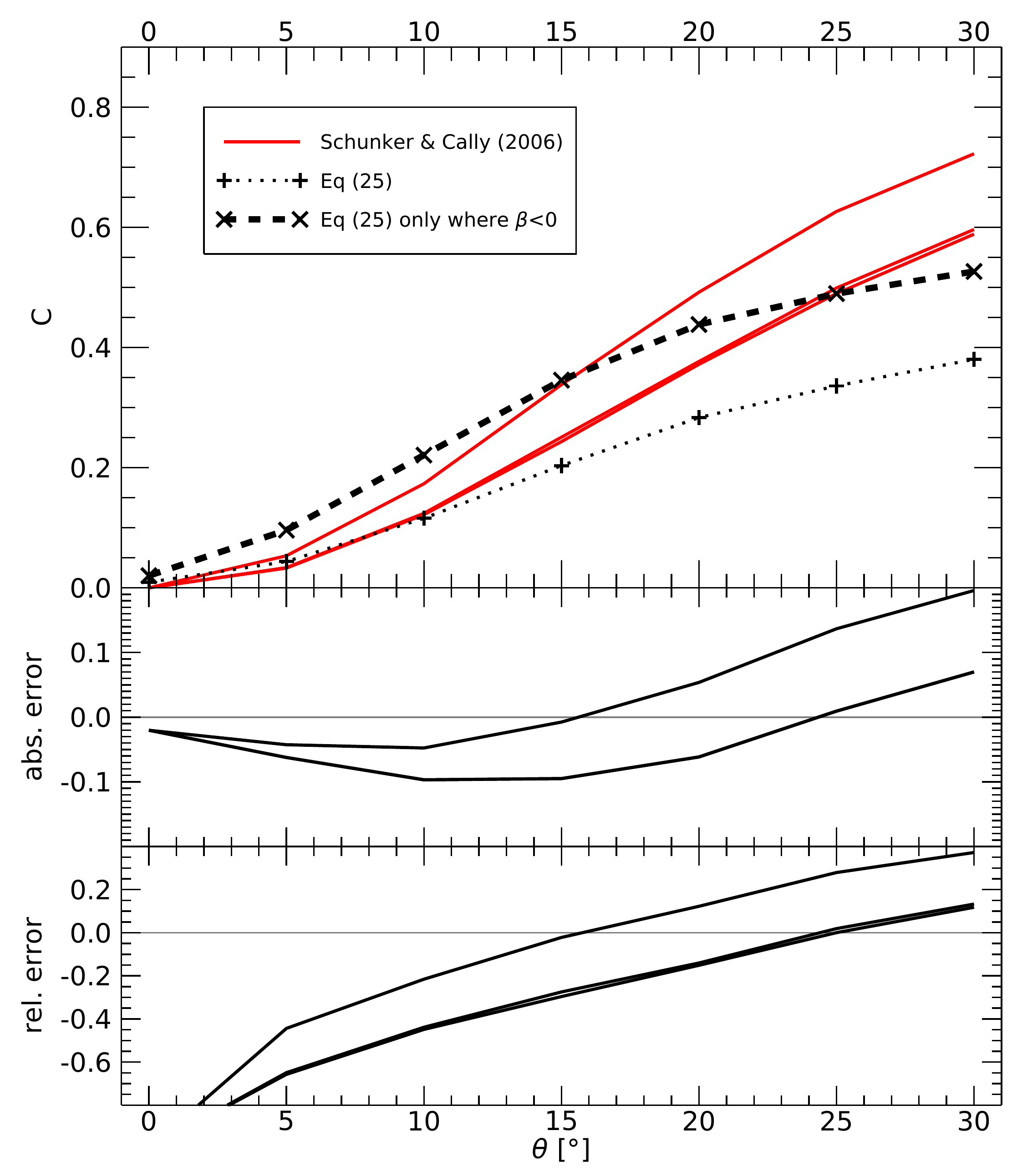}
\caption{\textit{Top:} Conversion coefficient describing mode conversion from acoustic to magnetic waves as a function of inclination angles. The full red lines are mode conversion according to \cite{schunker_cally_2006} (Equation \ref{eq:schunker_conversion}) in three different points within the loop exterior at the equipartition layer. The dotted black line is calculated according to Equation \ref{eq:conversion2} at $z=2$ Mm and averaged over the whole horizontal plane. The dashed black line is the same, but only considering fluxes in regions with $\beta <1$ (see text). \textit{Middle:} Absolute error of the red curves \citep{schunker_cally_2006} compared to the dashed curve (Eq. \ref{eq:conversion2} for $\beta<1$). \textit{Bottom:} Same, but relative error. \label{fig:conversion} }
\end{figure}

We calculate the conversion coefficient at three different points at the $c_A=c_s$ layer. All three points lie in the loop exterior below the transition region and are located at different arbitrary distances from the center of the loop exterior. Since the excited acoustic waves propagate vertically when below the transition region, we can assume that $k_z=k$ and $k_\perp=k \sin(\theta)$. Furthermore, we assume that $k=\omega / c_{\mathrm{eq}}$, where $\omega$ is our driver frequency and $c_{\mathrm{eq}}$ is the sound or Alfv\'{e}n speed at the equipartition layer. The resulting conversion coefficients are plotted in Fig. \ref{fig:conversion} (top) for various inclination angles in 5$\degr$ intervals as solid red lines. Two of those lines are more similar because the points in which they were calculated lie closer to each other than to the third point. The mode conversion clearly increases with increasing angle, as we expected. At a field inclination of about 15$\degr$ the curvature changes from positive to negative. 

To compare the outcome of Equation \ref{eq:schunker_conversion} with a reasonable quantity of our simulation results, we use the mean acoustic and mean magnetic energy flux defined by \cite{bray_loughhead_1974},
\begin{equation}
  \vec{F}_{ac}=\langle p_1 \vec{v}_1 \rangle,
\end{equation}
\begin{equation}
  \vec{F}_{mag}=\langle \vec{B}_1 \times (\vec{v}_1 \times \vec{B}_0) \rangle / \mu_0,
\end{equation}
where $p_1$, $\vec{v}_1$, and $\vec{B}_1$ are the perturbations of pressure, velocity, and magnetic field, respectively, and $\vec{B}_0$ is the equilibrium magnetic field. The (outer) brackets denote the average over time, however, for the following we also average over space. We now define a quantity
\begin{equation} \label{eq:conversion2}
  C_{f}=\frac{|\vec{F}_{mag}|}{|\vec{F}_{mag}|+|\vec{F}_{ac}|},
\end{equation}
which should tell us how much of the energy flux, which is initially fully acoustic, was converted into magnetic energy flux. Since we want to know how much our waves are dominated by which kind of energy flux, without taking into account the direction, we avoid positive and negative fluxes to be canceled out by taking the absolute value before averaging the fluxes. The value $C_{f}$ has the same properties as $C$ given that it is 1 for full conversion and 0 for no conversion. The dotted black line in Fig. \ref{fig:conversion} (top) shows $C_{f}$ averaged in time over half of a period at a height of 2 Mm, starting from the time when the first wave reaches that height, and averaged in space over the whole horizontal plane. The dashed black line shows the same, but only averaged in space over the areas where $\beta<1$. The latter line shows higher values, as there is more flux converted in the considered regions than in the rest of the plane because there the waves have to travel through the $\beta=1$ (and $c_A=c_s$) layer. That line is therefore more comparable to mode conversion in the three chosen points than the dotted line. Like before, we see more conversion from acoustic to magnetic waves with increasing magnetic field inclination $\theta$ and a change of curvature at around $\theta=15\degr$. 

By comparing $C_{f}$ (Equation \ref{eq:conversion2}) with $C$ (\cite{schunker_cally_2006}, Equation \ref{eq:schunker_conversion}) it is apparent that the curve of the latter conversion coefficient shows a higher inclination for increasing $\theta$ than the former. Fig. \ref{fig:conversion} shows the absolute (middle) and relative (bottom) errors of $C$ compared to $C_f$ (for $\beta<1$). Between  inclination angles of around $10\degr$ and $30\degr$ the relative error stays within $\pm 40 \%$. Below $10\degr$ the relative error is due to the small values of $C$ and $C_f$ much higher, however, the absolute error stays within the interval [-0.1,0].
In general, both conversion coefficients show the same qualitative behavior and are much more similar than we expected. This result is interesting because in our simulations we do not just have simple plane waves, except perhaps right above the driver at low heights along with some additional plane-like waves higher up. Instead, we mainly have waves that are modified by the cylindrical structure of the atmosphere, in particular sausage waves and kink waves. The local analysis around a point at the $c_A=c_s$ layer of \cite{schunker_cally_2006} allows the use of the simple analytic formula in Equation \ref{eq:schunker_conversion} for those cases as well.

We note that another, less general conversion coefficient was given by \cite{cally_2005}. It yields the same results as in Fig. \ref{fig:conversion} for $\theta \le 10\degr$, but deviates from Equation \ref{eq:schunker_conversion} for higher inclination angles, as the curve does not change its curvature. 

\section{Summary and conclusions} \label{sec:summary&conclusions}

In this paper we presented a simple method to calculate a 3D MHS equilibrium when a divergence-free magnetic field is given. We built an equilibrium model resembling the solar atmosphere from the photosphere to the lower corona, including four flux tubes of decreased Alfv\'{e}n speed and increased sound speed with the inclination $\theta$ from the vertical. This led to a tube-like plasma-$\beta=1$ layer with $\beta>1$ inside the tubes and everywhere in the bottom layers, and $\beta<1$ outside the tubes starting from a certain height. We then perturbed the plasma at the bottom with vertically polarized gravity-acoustic waves according to an analytical solution. We investigated the resulting waves by studying the behavior of the velocity perturbations parallel to the magnetic field, perpendicular to the magnetic iso-surfaces and azimuthal to these surfaces in a horizontal plane above the transition region. In addition, we studied the propagating velocity disturbances in two different lines (inside and outside the loop) along the magnetic field. By comparing our results with a simplified model of waves in a cylindrical structure, we could classify the waves appearing in our simulations. For the vertical case ($\theta=0\degr$), where the magnetic field and flux tubes are oriented along the driver polarization direction, we identified deformed fast sausage surface waves and slow sausage body waves. There might have additionally been a plane-like wave excited, which is difficult to distinguish from the fast sausage surface mode. For the inclined flux tubes and magnetic field, where the driver polarization now has a component perpendicular to the tubes, a fast kink surface wave is excited in conjunction with either a (deformed) fast sausage surface wave or a plane-like wave. Moreover, we also find slow sausage body waves.

In addition, we investigated the mode conversion from the initially acoustic waves to magnetic waves. We compared the outcome of a simple formula for a mode conversion coefficient by \cite{schunker_cally_2006} with the ratio of the magnetic energy flux to the sum of the magnetic and acoustic energy flux. We find that both methods give similar results with a maximum absolute error of 0.1 for inclination angles from $\theta=0\degr$ to $10\degr$ and a maximum relative error of 40\% for angles from $\theta=10\degr$ to $30\degr$. The deviation of the simple formula from the other method is remarkably small, given that we anticipated a large influence of the cross-field wave speed structuring. This validates the frequent use of the simple formula. We note, however, that the influence of a cutoff region was not tested in the present work.

According to our simulations, vertical gravity-acoustic waves from the photosphere are converted to waves with partial magnetic properties in areas with flux tubes (especially in between the tubes), if the magnetic field lines and the flux tubes are inclined from the vertical. In that case the initially vertically propagating plane waves changed direction to propagate along the magnetic field above the transition region and were transformed into kink and sausage modes. In the case with vertical magnetic field and flux tubes, we only observed sausage waves without any significant magnetic wave properties.

There are some important limitations of this work we would like to mention. First, because of our model containing straight flux tubes, the magnetic field does not change along the loops to satisfy $\nabla \cdot \vec{B}=0$. However, this leads to an unusually high magnetic field strength of 300 G in the corona. In addition, there are regions in our model with $\beta>1$ in the corona, which is a much higher value than expected for that height \citep[see, e.g., ][]{gary_2001}. Realistic flux tubes are expected to strongly expand between photosphere and lower corona. Such a model would not only allow the magnetic field to decrease with height, but the expansion would also affect waves guided along the flux tubes. We assume the biggest change would be that the wave fronts are refracted along the field lines and would broaden, which would lead to damping of the waves. This was also mentioned in the results of \cite{mumford_etal_2015}. In fact, we assume the simple geometry of our flux tubes to be the main reason why the damping of the waves in our simulations is far smaller than in the observations of \cite{grant_etal_2015}. Despite these drawbacks we decided to first study straight flux tubes as they simplify the analysis of the excited waves. A similar study with expanding flux tubes is currently in progress.

A second limitation of this study are the high frequency and low amplitude of our driver. While we do not think that a higher amplitude would change the core of our results, we assume that a lower frequency of the driver with a realistic period of about five minutes leads to less waves being transmitted into the corona. This is because of the acoustic cutoff region, which prohibits the propagation of acoustic waves below the cutoff frequency. Instead, many waves would be reflected from that region. Since the wavelength in the current study is already much larger than the flux tube radius, we do not expect other big changes by decreasing the frequency. We plan to use a driver period of approximately five minutes in future work.

\begin{acknowledgements}
We would like to thank Paul Cally for helpful comments and suggestions. Furthermore, we would like to thank the referee for helping us to improve the manuscript. This project has received funding from the European Research Council (ERC) under the European Union's Horizon 2020 research and innovation programme (grant agreement No. 724326).
\end{acknowledgements}

\bibliographystyle{aa}
\bibliography{../sources}

\begin{thebibliography}{52}
\expandafter\ifx\csname natexlab\endcsname\relax\def\natexlab#1{#1}\fi

\bibitem[{{Anfinogentov} {et~al.}(2013){Anfinogentov}, {Nistic{\`o}}, \&
  {Nakariakov}}]{anfinogentov_etal_2013}
{Anfinogentov}, S., {Nistic{\`o}}, G., \& {Nakariakov}, V.~M. 2013, \aap, 560,
  A107

\bibitem[{{Arregui}(2017)}]{arregui_2017}
{Arregui}, I. 2017, in SOLARNET IV: The Physics of the Sun from the Interior to
  the Outer Atmosphere, 59

\bibitem[{Aschwanden(2006)}]{aschwanden}
Aschwanden, M. 2006, Physics of the Solar Corona: An Introduction with Problems
  and Solutions, second edition edn., Springer Praxis Books / Astronomy and
  Planetary Sciences (Springer)

\bibitem[{{Bogdan} {et~al.}(1996){Bogdan}, {Hindman}, {Cally}, \&
  {Charbonneau}}]{bogdan_etal_1996}
{Bogdan}, T.~J., {Hindman}, B.~W., {Cally}, P.~S., \& {Charbonneau}, P. 1996,
  \apj, 465, 406

\bibitem[{{Bray} \& {Loughhead}(1974)}]{bray_loughhead_1974}
{Bray}, R.~J. \& {Loughhead}, R.~E. 1974, {The solar chromosphere}

\bibitem[{{Cally}(2005)}]{cally_2005}
{Cally}, P.~S. 2005, \mnras, 358, 353

\bibitem[{{Cally}(2017)}]{cally_2017}
{Cally}, P.~S. 2017, \mnras, 466, 413

\bibitem[{{Centeno} {et~al.}(2006){Centeno}, {Collados}, \& {Trujillo
  Bueno}}]{centeno_etal_2006}
{Centeno}, R., {Collados}, M., \& {Trujillo Bueno}, J. 2006, \apj, 640, 1153

\bibitem[{{De Pontieu} {et~al.}(2005){De Pontieu}, {Erd{\'{e}}lyi}, \& {De
  Moortel}}]{de_pontieu_etal_2005}
{De Pontieu}, B., {Erd{\'{e}}lyi}, R., \& {De Moortel}, I. 2005, The
  Astrophysical Journal, 624, L61

\bibitem[{{De Pontieu} {et~al.}(2004){De Pontieu}, {Erd{\'e}lyi}, \&
  {James}}]{de_pontieu_etal_2004}
{De Pontieu}, B., {Erd{\'e}lyi}, R., \& {James}, S.~P. 2004, \nat, 430, 536

\bibitem[{{De Pontieu} {et~al.}(2007){De Pontieu}, {McIntosh}, {Carlsson},
  {Hansteen}, {Tarbell}, {Schrijver}, {Title}, {Shine}, {Tsuneta}, {Katsukawa},
  {Ichimoto}, {Suematsu}, {Shimizu}, \& {Nagata}}]{de_pontieu_etal_2007}
{De Pontieu}, B., {McIntosh}, S.~W., {Carlsson}, M., {et~al.} 2007, Science,
  318, 1574

\bibitem[{{de Wijn} {et~al.}(2009){de Wijn}, {McIntosh}, \& {De
  Pontieu}}]{de_wijn_etal_2009}
{de Wijn}, A.~G., {McIntosh}, S.~W., \& {De Pontieu}, B. 2009, \apjl, 702, L168

\bibitem[{{Edl{\'e}n}(1943)}]{edlen_1943}
{Edl{\'e}n}, B. 1943, \zap, 22, 30

\bibitem[{{Fedun} {et~al.}(2011){Fedun}, {Shelyag}, \&
  {Erd{\'e}lyi}}]{fedun_etal_2011}
{Fedun}, V., {Shelyag}, S., \& {Erd{\'e}lyi}, R. 2011, \apj, 727, 17

\bibitem[{{Felipe} {et~al.}(2010){Felipe}, {Khomenko}, \&
  {Collados}}]{felipe_etal_2010}
{Felipe}, T., {Khomenko}, E., \& {Collados}, M. 2010, \apj, 719, 357

\bibitem[{{Gary}(2001)}]{gary_2001}
{Gary}, G.~A. 2001, \solphys, 203, 71

\bibitem[{{Gascoyne} {et~al.}(2014){Gascoyne}, {Jain}, \&
  {Hindman}}]{gascoyne_etal_2014}
{Gascoyne}, A., {Jain}, R., \& {Hindman}, B.~W. 2014, \apj, 789, 109

\bibitem[{{Grant} {et~al.}(2015){Grant}, {Jess}, {Moreels}, {Morton},
  {Christian}, {Giagkiozis}, {Verth}, {Fedun}, {Keys}, {Van Doorsselaere}, \&
  {Erd{\'e}lyi}}]{grant_etal_2015}
{Grant}, S.~D.~T., {Jess}, D.~B., {Moreels}, M.~G., {et~al.} 2015, \apj, 806,
  132

\bibitem[{{Griffiths} {et~al.}(2018){Griffiths}, {Fedun}, {Erd{\'e}lyi}, \&
  {Zheng}}]{griffiths_2018}
{Griffiths}, M.~K., {Fedun}, V., {Erd{\'e}lyi}, R., \& {Zheng}, R. 2018,
  Advances in Space Research, 61, 720

\bibitem[{{Guo} {et~al.}(2019){Guo}, {Van Doorsselaere}, {Karampelas}, {Li},
  {Antolin}, \& {De Moortel}}]{guo_etal_2019}
{Guo}, M., {Van Doorsselaere}, T., {Karampelas}, K., {et~al.} 2019, \apj, 870,
  55

\bibitem[{{Hansen} \& {Cally}(2009)}]{hansen_cally_2009}
{Hansen}, S.~C. \& {Cally}, P.~S. 2009, \solphys, 255, 193

\bibitem[{{Hindman} \& {Jain}(2008)}]{hindman_jain_2008}
{Hindman}, B.~W. \& {Jain}, R. 2008, \apj, 677, 769

\bibitem[{{Karampelas} \& {Van
  Doorsselaere}(2018)}]{karampelas_vandoorsselaere_2018}
{Karampelas}, K. \& {Van Doorsselaere}, T. 2018, \aap, 610, L9

\bibitem[{{Karampelas} {et~al.}(2017){Karampelas}, {Van Doorsselaere}, \&
  {Antolin}}]{karampelas_etal_2017}
{Karampelas}, K., {Van Doorsselaere}, T., \& {Antolin}, P. 2017, \aap, 604,
  A130

\bibitem[{{Karampelas} {et~al.}(2019){Karampelas}, {Van Doorsselaere}, \&
  {Guo}}]{karampelas_etal_2019}
{Karampelas}, K., {Van Doorsselaere}, T., \& {Guo}, M. 2019

\bibitem[{{Khomenko} \& {Collados}(2006)}]{khomenko_collados_2006}
{Khomenko}, E. \& {Collados}, M. 2006, \apj, 653, 739

\bibitem[{{Khomenko} {et~al.}(2008){Khomenko}, {Collados}, \&
  {Felipe}}]{khomenko_etal_2008}
{Khomenko}, E., {Collados}, M., \& {Felipe}, T. 2008, \solphys, 251, 589

\bibitem[{{Khomenko} {et~al.}(2018){Khomenko}, {Vitas}, {Collados}, \& {de
  Vicente}}]{khomenko_etal_2018}
{Khomenko}, E., {Vitas}, N., {Collados}, M., \& {de Vicente}, A. 2018

\bibitem[{{Krishna Prasad} {et~al.}(2012){Krishna Prasad}, {Banerjee}, {Van
  Doorsselaere}, \& {Singh}}]{krishna_etal_2012}
{Krishna Prasad}, S., {Banerjee}, D., {Van Doorsselaere}, T., \& {Singh}, J.
  2012, \aap, 546, A50

\bibitem[{{Marsh} \& {Walsh}(2006)}]{marsh_walsh_2006}
{Marsh}, M.~S. \& {Walsh}, R.~W. 2006, \apj, 643, 540

\bibitem[{{Mihalas} \& {Mihalas}(1984)}]{mihalas_mihalas_1984}
{Mihalas}, D. \& {Mihalas}, B.~W. 1984, {Foundations of radiation
  hydrodynamics}

\bibitem[{{Moreels} \& {Van Doorsselaere}(2013)}]{moreels_vandoorsselaere_2013}
{Moreels}, M.~G. \& {Van Doorsselaere}, T. 2013, \aap, 551, A137

\bibitem[{{Morton} {et~al.}(2016){Morton}, {Tomczyk}, \&
  {Pinto}}]{morton_etal_2016}
{Morton}, R.~J., {Tomczyk}, S., \& {Pinto}, R.~F. 2016, \apj, 828, 89

\bibitem[{{Morton} {et~al.}(2012){Morton}, {Verth}, {Jess}, {Kuridze},
  {Ruderman}, {Mathioudakis}, \& {Erd{\'e}lyi}}]{morton_etal_2012}
{Morton}, R.~J., {Verth}, G., {Jess}, D.~B., {et~al.} 2012, Nature
  Communications, 3, 1315

\bibitem[{{Morton} {et~al.}(2019){Morton}, {Weberg}, \&
  {McLaughlin}}]{morton_etal_2019}
{Morton}, R.~J., {Weberg}, M.~J., \& {McLaughlin}, J.~A. 2019, Nature Astronomy
  

\bibitem[{{Mumford} {et~al.}(2015){Mumford}, {Fedun}, \&
  {Erd{\'e}lyi}}]{mumford_etal_2015}
{Mumford}, S.~J., {Fedun}, V., \& {Erd{\'e}lyi}, R. 2015, \apj, 799, 6

\bibitem[{{Nakariakov} {et~al.}(2016){Nakariakov}, {Anfinogentov},
  {Nistic{\`o}}, \& {Lee}}]{nakariakov_etal_2016}
{Nakariakov}, V.~M., {Anfinogentov}, S.~A., {Nistic{\`o}}, G., \& {Lee}, D.-H.
  2016, \aap, 591, L5

\bibitem[{{Nistic{\`o}} {et~al.}(2013){Nistic{\`o}}, {Nakariakov}, \&
  {Verwichte}}]{nistico_etal_2013}
{Nistic{\`o}}, G., {Nakariakov}, V.~M., \& {Verwichte}, E. 2013, \aap, 552, A57

\bibitem[{{Pagano} \& {De Moortel}(2019)}]{pagano_de_moortel_2019}
{Pagano}, P. \& {De Moortel}, I. 2019

\bibitem[{{Parnell} \& {De Moortel}(2012)}]{parnell_demoortel_2012}
{Parnell}, C.~E. \& {De Moortel}, I. 2012, Philosophical Transactions of the
  Royal Society of London Series A, 370, 3217

\bibitem[{{Pascoe} {et~al.}(2007){Pascoe}, {Nakariakov}, \&
  {Arber}}]{pascoe_etal_2007}
{Pascoe}, D.~J., {Nakariakov}, V.~M., \& {Arber}, T.~D. 2007, \aap, 461, 1149

\bibitem[{Prasad {et~al.}(2015)Prasad, Jess, \& Khomenko}]{prasad_etal_2015}
Prasad, S.~K., Jess, D.~B., \& Khomenko, E. 2015, The Astrophysical Journal,
  812, L15

\bibitem[{Priest(2014)}]{priest_2014}
Priest, E. 2014, Magnetohydrodynamics of the Sun (Cambridge University Press)

\bibitem[{{Raes} {et~al.}(2017){Raes}, {Van Doorsselaere}, {Baes}, \&
  {Wright}}]{raes_etal_2017}
{Raes}, J.~O., {Van Doorsselaere}, T., {Baes}, M., \& {Wright}, A.~N. 2017,
  \aap, 602, A75

\bibitem[{{Santamaria} {et~al.}(2015){Santamaria}, {Khomenko}, \&
  {Collados}}]{santamaria_etal_2015}
{Santamaria}, I.~C., {Khomenko}, E., \& {Collados}, M. 2015, \aap, 577, A70

\bibitem[{{Schunker} \& {Cally}(2006)}]{schunker_cally_2006}
{Schunker}, H. \& {Cally}, P.~S. 2006, \mnras, 372, 551

\bibitem[{{Tarr} {et~al.}(2017){Tarr}, {Linton}, \& {Leake}}]{tarr_etal_2017}
{Tarr}, L.~A., {Linton}, M., \& {Leake}, J. 2017, \apj, 837, 94

\bibitem[{{Tomczyk} \& {McIntosh}(2009)}]{tomczyk_mcintosh_2009}
{Tomczyk}, S. \& {McIntosh}, S.~W. 2009, \apj, 697, 1384

\bibitem[{{Tomczyk} {et~al.}(2007){Tomczyk}, {McIntosh}, {Keil}, {Judge},
  {Schad}, {Seeley}, \& {Edmondson}}]{tomczyk_etal_2007}
{Tomczyk}, S., {McIntosh}, S.~W., {Keil}, S.~L., {et~al.} 2007, Science, 317,
  1192

\bibitem[{{Verwichte} {et~al.}(2005){Verwichte}, {Nakariakov}, \&
  {Cooper}}]{verwichte_etal_2005}
{Verwichte}, E., {Nakariakov}, V.~M., \& {Cooper}, F.~C. 2005, \aap, 430, L65

\bibitem[{{Wang} {et~al.}(2012){Wang}, {Ofman}, {Davila}, \&
  {Su}}]{wang_etal_2012}
{Wang}, T., {Ofman}, L., {Davila}, J.~M., \& {Su}, Y. 2012, \apjl, 751, L27

\bibitem[{{Zhao} {et~al.}(2016){Zhao}, {Felipe}, {Chen}, \&
  {Khomenko}}]{zhao_etal_2016}
{Zhao}, J., {Felipe}, T., {Chen}, R., \& {Khomenko}, E. 2016, \apjl, 830, L17

\end{thebibliography}

\end{document}